\documentclass[%
 reprint,
nobibnotes,
 amsmath,amssymb,
 aps,
 prb,
]{revtex4-1}
\usepackage{multirow}
\usepackage[utf8]{inputenc}          
\usepackage[T1]{fontenc}             
\usepackage{graphicx}                
\usepackage{epstopdf}                
\usepackage{bm}                      
\usepackage[version=4]{mhchem}       
\usepackage{booktabs}                
\usepackage{afterpage}              
\usepackage{rotating}
\usepackage{appendix}
\usepackage{setspace}               
\usepackage{soul}                   
\usepackage{xcolor}                 
\usepackage[colorlinks=true,       
            linkcolor=blue,
            citecolor=blue,
            urlcolor=blue]{hyperref} 

\newcommand{\angstrom}{\textup{\AA}}
\expandafter\ifx\csname package@font\endcsname\relax\else
 \expandafter\expandafter
 \expandafter\usepackage
 \expandafter\expandafter
 \expandafter{\csname package@font\endcsname}%
\fi
\begin{document}

\setstretch{1.0}
\title{Molecular Dynamics Simulations of $\gamma$-Belite(010)-Water Interfaces with High-Dimensional Neural Network Potentials}
\author{Bernadeta Prus}
\author{J\"{o}rg Behler}
\email{joerg.behler@rub.de}
\affiliation{Lehrstuhl f\"ur Theoretische Chemie II, Ruhr-Universit\"at Bochum, 44780 Bochum, Germany, and Atomistic Simulations, Research Center Chemical Sciences and Sustainability, Research Alliance Ruhr, 44780 Bochum, Germany}

\date{\today}

\begin{abstract}
Belite -- dicalcium silicate Ca$_2$SiO$_4$ -- is a main constituent of low-carbon cement. In this work, we study several terminations of the (010) surface of $\gamma$-belite, its most stable polymorph, by molecular dynamics simulations. The energies and forces are provided by a high-dimensional neural network potential trained to density functional theory data. Water can interact in molecular form as well as dissociatively with the investigated interfaces, and the degree of dissociation is determined primarily by the protonation of SiO$_4$ groups accessible at the surface. A major part of the simultaneously formed hydroxide ions is adsorbed at surface calcium atoms, whose octahedral coordination spheres are completed by additional water molecules. The T3 termination, which is most stable in vacuum, shows only little reactivity in water. For the only slightly less stable T2 termination, however, two distinct types of surface defects are observed. The type I defect is even stable in vacuum and leads to a reconstruction of the entire surface, while the type II defect is only found in the presence of water. Overall, our results suggest that a variety of structures may be formed at the Ca$_2$SiO$_4$(010) surface, which are stabilized in the presence of water.
\end{abstract}

\maketitle

\section{Introduction}\label{sec:introduction}

Concrete, made by the reaction of water with cement clinker, is the most widespread building material world-wide. The most common starting material for concrete is Portland cement containing four main components: tricalcium silicate (alite, Ca$_3$SiO$_5$), dicalcium silicate (belite, Ca$_2$SiO$_4$), tricalcium aluminate (Ca$_3$Al$_2$O$_6$), and tetracalcium aluminoferrite (Ca$_4$Al$_2$Fe$_2$O$_{10}$)~\cite{P7224,P6566}. The properties of concrete are determined by the hydration processes of these minerals in contact with water, which result, e.g., in structurally complex calcium silicate hydrates.

A substantial amount of CO$_2$ is generated in the production of cement clinker, resulting from the use of fossil fuels for calcinating the raw materials at high temperatures as well as the release of chemically bound CO$_2$ from the heated limestone, and in recent years this has become a major concern~\cite{P7146,P7224,P6729}. Consequently, the search for alternative cement mixtures with reduced greenhouse gas emissions has received increasing attention~\cite{P6870,P6871}. A promising route to a lower CO$_2$ footprint of concrete is to increase the contents of belite in cement clinker~\cite{P6870}, which contains less calcium, i.e., requires less CaCO$_3$ starting material, and can be produced at lower temperatures. 

Belite exists in several polymorphs ($\alpha$, $\beta$, and $\gamma$), among which $\gamma$-belite is most stable at ambient conditions~\cite{P6873}. Unfortunately, in contrast to other highly reactive components of Portland cement, $\gamma$-belite shows only a very low reactivity with respect to water~\cite{P6869,P6878}, resulting in a slow hydration process. Understanding this low reactivity of belite in detail is an important condition for enhancing the activity of its hydration process~\cite{P6871,P7147,P6889}. 

Belite and its interfaces have been studied by computer simulations using a wide range of methods~\cite{P6523,P6755,P7223}. Many properties of the bulk polymorphs and of its surfaces have been investigated by accurate density-functional theory (DFT) calculations~\cite{P6878,P6880,P6881,P6756,P7145}. However, the high computational costs of DFT-based \emph{ab initio} molecular dynamics (MD) simulations do not allow to perform systematic studies of its interfaces with a realistic liquid phase of water, and so far only small model systems could be investigated on short time scales~\cite{P6590,P7156}. 

To address more complex systems, which are particularly important to describe the amorphous phases encountered in the hydration process, very efficient classical force fields have been extensively used~\cite{P6655,P6657,P6753,P6764,P6875,P7153}. Some important force fields that have been applied in concrete chemistry are, e.g., ReaxFF~\cite{P0928,P7148,P6749,P6888}, ClayFF~\cite{P7149,P6748,P7115}, and cemff~\cite{P6656}. However, while employing these methods much larger systems can be studied at long time scales, the accuracy is necessarily restricted by the underlying approximations.

To overcome these limiations, in recent years a lot of effort has been put in the development of a new generation of interatomic potentials based on machine learning techniques. Once trained to a reference set of electronic structure data, typically employing DFT, these machine learning potentials (MLP) allow to compute the energies and forces of large systems with almost the accuracy of quantum mechanics at much reduced costs~\cite{P4885,P5673,P5793,P6102,P6121,P6112}. Many applications of MLPs for a wide range of systems in chemistry and materials science have been reported to date. Among these studies, also some applications of MLPs to cement minerals have been published addressing, e.g., the properties of tobermorite and calcium silicate hydrates~\cite{P6752,P7114,P7137,P7150,P7151,P7152} but also the the dissolution of $\beta$-belite in water~\cite{P6582}. Moreover, recently a machine learning potential was used to search for new polymorphs of belite~\cite{P6769}. Still, the aqueous interfaces of the most stable polymorph $\gamma$-belite, which are of high interest for improving the activity of the cement hydration process, have not been studied so far using MLPs. 

In the present work, we fill this gap by performing large-scale MD simulations of the low-energy terminations of the most stable $\gamma$-belite(010) surface in contact with bulk liquid water. The energies and forces for the simulations are provided by a high-dimensional neural network potential (HDNNP)~\cite{P1174}, an efficient type of MLP that has been frequently applied to study many different types of solid-liquid interfaces~\cite{P6721,P7208}. After training and validating the HDNNP using DFT data, we investigate the structure and reactivity of the water molecules at a series of interface in detail. Moreover, we combine DFT calculations and HDNNP-driven MD simulations to further investigate and characterize some structural changes of the systems emerging in the simulations. 

\section{High-Dimensional Neural Network Potentials}\label{sec:methods}

In the present work, we use second-generation (2G) HDNNPs to represent the energy and atomic forces with quantum mechanical accuracy~\cite{P1174,P6018}. In 2G-HDNNPs, the potential energy $E$ of a system consisting of $N_{\mathrm{atoms}}$ atoms is constructed as a sum of atomic energies $E_i$,
\begin{equation}
E=\sum_{i=1}^{N_{\mathrm{atoms}}}E_i \quad .
\label{eq:energy_HDNNP}
\end{equation}
The atomic energies depend on the neighboring atoms within a spatial cutoff radius, which is typically between 5 and 10 \AA{}. For each atom in the system, the positions of its neighbors inside the cutoff are encoded as a vector of atom-centered symmetry functions (ACSF)~\cite{P2882}, which ensure the invariance of the potential energy with respect to translation and rotation of the system as well as permutation of atoms of the same element. The ACSF vectors serve as input to element-specific atomic feed-forward neural networks, which are trained by optimizing the neural network weight parameters to reproduce a reference set of energies and forces from electronic structure calculations. HDNNPs are very efficient, and once trained, they can be applied to systems containing thousands of atoms to perform large-scale molecular dynamics simulations on nanosecond timescales. The required forces and stress tensors can be computed analytically as the corresponding energy derivatives.

\section{Computational Details}\label{sec:computational}

\subsection{Slab setup in vacuum} 

\begin{figure}
    \centering
    \includegraphics[width=0.7\linewidth]{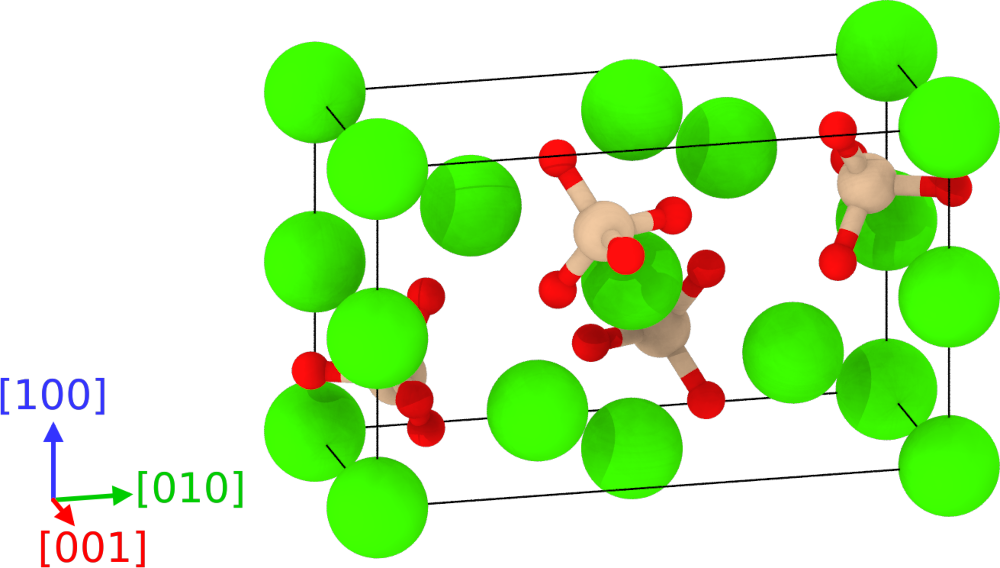}
    \caption{DFT-optimized orthorhombic unit cell of $\gamma$-belite containing four formula units of Ca$_2$SiO$_4$. Atoms are represented as spheres in colors corresponding to the elements: Ca - green, Si - beige, O - red.}
    \label{fig:bulk_gamma}
\end{figure}

\begin{figure*}
\includegraphics[width=1.00\linewidth]{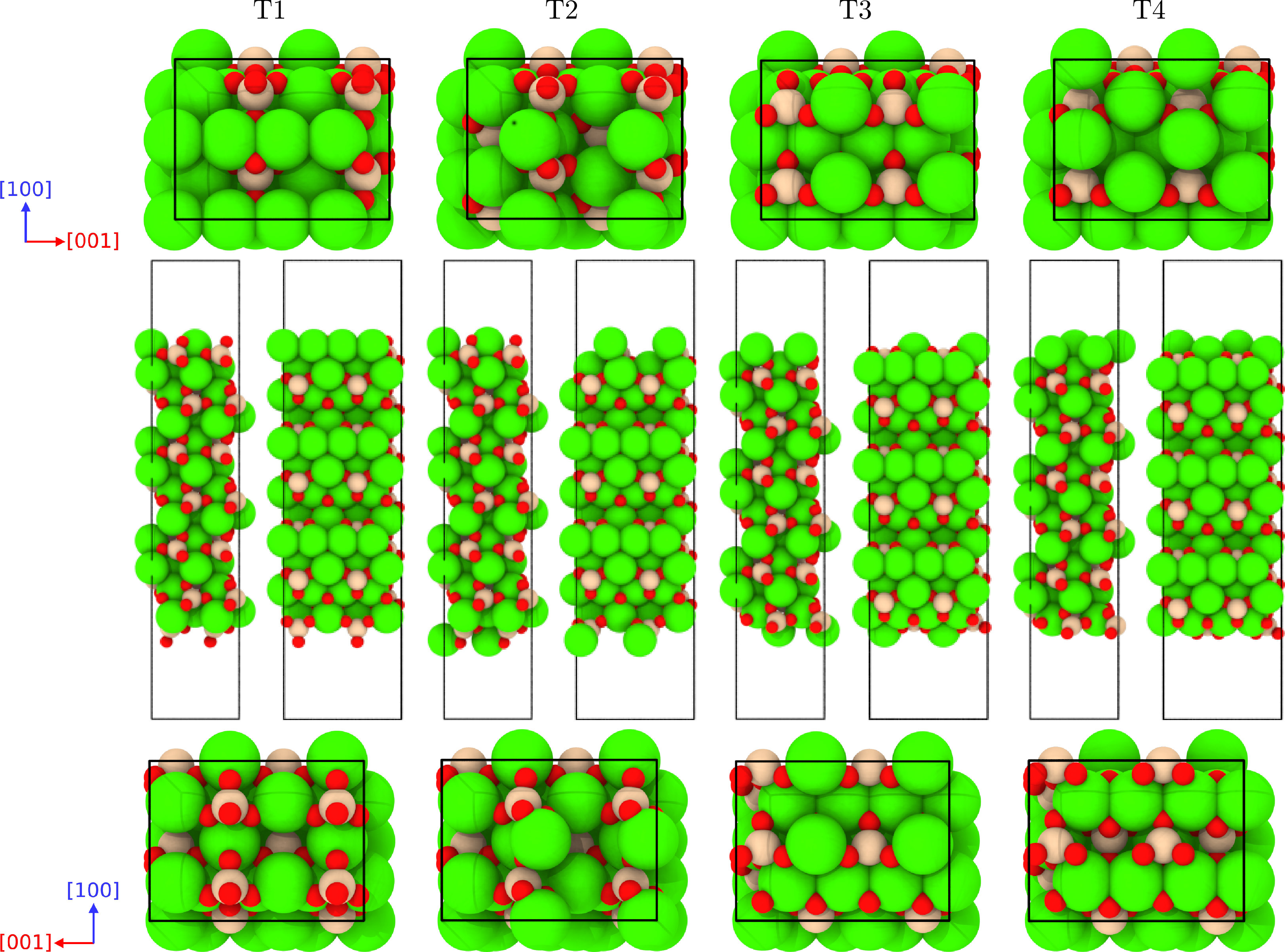}
\caption{DFT-optimized (2$\times$2) supercells of the (010) surfaces of $\gamma$-belite with different terminations. The corresponding cleavage energies are given in Table~\ref{table:cleavage_energy}. The first row presents top views, the second row presents side views, alternating between the [100] and [001] directions, while the last row shows bottom views. The top and bottom surfaces are identical for the non-polar T2 and T3 terminations, while they are different for the polar T1 and T4 terminated slabs.}
\label{fig:surfaces}
\end{figure*}

The DFT-optimized orthorhombic bulk unit cell of $\gamma$-belite is shown in Fig.~\ref{fig:bulk_gamma}.
In the present work, we focus on the (010) surface, which has been reported to be most stable in previous work~\cite{P6881}.
Depending on the cutting position, for this surface there are four different possible surface terminations T1-T4 (Fig.~\ref{fig:surfaces}). For computing their cleavage energies, slabs were prepared based on (1$\times$3$\times$1) supercells of the bulk structure resulting in a slab thickness of about 34~\AA{} with 20~\AA{} of vacuum in between. When cutting the surfaces, the SiO$_4$ tetrahedra were preserved to avoid high energy structures resulting from the cleavage of Si-O bonds~\cite{P6582}. During geometry optimizations, the central part of the slab equivalent to one unit cell of bulk $\gamma$-belite was frozen while relaxing the top and bottom parts each having a thickness of about 11~\AA{}. 

As shown in Fig.~\ref{fig:surfaces}, the DFT-optimized slabs with T2 and T3 terminations are symmetric, i.e., they have the same top and bottom surface geometries. Contrarily, the polar T1 and T4 slabs exhibit different structures at both sides, where one side is terminated by calcium ions and the other by SiO$_4$ tetrahedra. This asymmetry is required to maintain the overall stoichiometry of the slab preventing charged systems. Therefore, in general only averaged cleavage energies $\gamma$ of both surface geometries can be calculated according to
\begin{equation}
\gamma=\frac{1}{2A}(E_\text{slab}-nE_\text{bulk}) \quad ,
\label{eq:cleavage_energy}
\end{equation}
where $A$ is the surface area, $E_\text{slab}$ is the DFT energy of the relaxed slab, $n$ is the number of bulk unit cells used for the construction of the slab, and $E_\text{bulk}$ is the energy of the DFT-optimized bulk unit cell. 

The obtained cleavage energies are compiled in Table~\ref{table:cleavage_energy}. The value of 1.14~J/m$^{2}$ for the most stable T3 termination is in excellent agreement with 1.13~J/m$^{2}$ reported in previous work \cite{P6590}. The second most stable termination is T2 with a cleavage energy of about 1.50~J/m$^{2}$, whereas the clearly less stable T1 and T4 terminations exhibit more than two times higher cleavage energies than T3. Based on these findings, in this work we will focus on the two most stable terminations, T2 and T3, of the $\gamma$-belite (010) surface. 

In the bulk structure of $\gamma$-belite, each calcium atom is octahedrally coordinated by six oxygen atoms and each oxygen atom is tetrahedrally coordinated by three calcium atoms and one silicon atom. At the surfaces of the slabs, some of these neighbors are missing. The calcium coordination numbers for the four terminations are presented in Table~\ref{table:cleavage_energy}. Due to the different top and bottom surfaces, the polar T1 and T4 slabs are characterized by different coordination numbers on both sides. Calcium atoms at the bottom surface and oxygen atoms at the top surface of T4 are fully coordinated, while all other surfaces show varying degrees of undercoordinated oxygen and calcium atoms. The most stable terminations T2 and T3 chosen for further analysis exhibit a low Ca coordination number of 3, which in case of cement minerals is often associated with higher reactivity due to the possible coordination of calcium by the oxygen atoms of water molecules~\cite{Qi_2021} or hydroxide ions. Additionally, the non-polar T2 and T3 terminations exhibit both, under-coordinated calcium and oxygen atoms, such that these surfaces require proton as well as hydroxide and/or water adsorption for full saturation. For the top surface of T1 and both surfaces of T2, even the second layer of calcium atoms is not fully coordinated by oxygen.

\setlength{\tabcolsep}{2pt}
\begin{table*}[htbp]
\begin{center}
\caption{DFT cleavage energies averaged over both surfaces for different terminations of $\gamma$-belite (010) surfaces, and calcium coordination numbers with respect to oxygen atoms.
Values in brackets refer to Ca atoms in the second layer that are not fully coordinated in the bulk-truncated structure. 
The surfaces are visualized in Fig.~\ref{fig:surfaces}. 
}
\label{table:cleavage_energy}
\setlength{\tabcolsep}{12pt}
\begin{tabular}{lcccc} 
 \toprule	
\multirow{2}{*}{Parameter:} & \multicolumn{4}{l}{Termination:} \\
\cline{2-5}
  & T1 & T2 & T3 & T4 \\ 
    \midrule
 Cleavage energy (J/m$^{2}$) & 2.85 & 1.50 & 1.14 & 2.72 \\  
 Top surface Ca coordination number(s) & 3(5) & 3(5) & 3 & 3\\  
 Bottom surface  Ca coordination number(s) & 5/6 & 3(5) & 3 & 6 \\
 \bottomrule	
\end{tabular}
\end{center}
\end{table*}

\subsection{Construction of the HDNNP}

The HDNNPs were trained employing the \textit{RuNNer} code~\cite{Behler_2015,Behler_2017} using a DFT dataset obtained by active learning (cf. Section D in the Supporting Information). The local atomic environments with a cutoff radius of 6~\AA{} are described by 188, 191, 167, and 165 ACSFs~\cite{P2882} for hydrogen, oxygen, calcium, and silicon atoms, respectively. The parameters of these ACSFs are compiled in the Supporting Information in Section E. The atomic neural networks of all elements have the same architecture and consist of two hidden layers containing 15 neurons each. The hyperbolic tangent has been used as activation function in the hidden layers, while a linear function has been employed for the output nodes. The weight parameters have been optimized using energies and forces of approximately 90~\% of the available reference data employing a global adaptive Kalman filter~\cite{P1308}, while 10~\% of the data have been used as an independent test set. Further details about the general training and validation procedure of HDNNPs can be found in Refs.~\citenum{P4444} and~\citenum{P6548}. 

\subsection{Trajectory Analysis}

A description of the settings of the Molecular Dynamics simulations carried out in this work for the equilibration of the system and the production runs is given in Section B in the Supplementary Information.
For a deeper analysis, different water species like H$_2$O molecules and hydroxide ions in the system were identified in the MD trajectories by assigning each H atom to the closest O atom.  The following groups of species can be distinguished: species adsorbed at the surface are labeled with ``*'' (adsorbed water molecules H$_2$O*, adsorbed hydroxide ions O*H$^-$), species containing surface oxygen atoms are assigned a subscript ``s'' (unprotonated surface oxygen ions O$_\text{s}^{2-}$, protonated surface oxygen ions O$_\text{s}$H$^-$), and non-absorbed species are indicated by "$^{\dagger}$" (fully solvated hydroxide ions O$^{\dagger}$H$^-$, bulk water molecules H$_2$O$^{\dagger}$). The O-Ca distance threshold for adsorption was defined as 2.9~\AA{}, based on the decay of the first peak in the radial distribution function of bulk $\gamma$-belite (see Supplementary Information Fig.~S1). The analyses of all species were performed using the lionanalysis code~\cite{lion}. In the following discussion, we will adopt the notation that for the slab surface layer number 1 corresponds to the outermost Ca atom layer, while Ca layer 2 is the first subsurface layer. Contrarily, for the adsorbed oxygen species we follow the usual convention that the first layer refers to the atoms closest to the slab, and the numbering of layers increases with separation from the surface. 

\section{Results and discussion}

\subsection{High-dimensional neural network potential}

The HDNNP is based on a DFT reference dataset obtained with the RPBE functional~\cite{P0120} and D3 dispersion corrections~\cite{P3112}, which consists of 3,361 bulk $\gamma$-belite, 6,756 bulk liquid water, and 14,267 interface structures covering the T2 and T3 terminations of the (010) surface. This dataset of overall 24,384 structures, which contain between 24 and 1620 atoms, covers 4,817,928 atomic environments providing 24,384 total energies and 14,453,784 force components. More detailed information about the construction of the reference data set and its composition is given in the Supplementary Information Section D. 

For training the HDNNP, the dataset has been randomly split into a training set consisting of 21,938 structures and an independent test set of 2,446 structures. The root mean squared errors (RMSEs) of the energies are 0.627 meV/atom for the training set and 0.654 meV/atom for the test set, while the RMSEs of the force components are 36.4 meV/Bohr and 36.2 meV/Bohr for the training and the test set, respectively. Further details about the HDNNP and its validation are presented in the Supplementary Information Sections E-H.

\subsection{\texorpdfstring{Bulk $\gamma$-dicalcium silicate}{Bulk gamma-dicalcium silicate}}

As a first step, we have investigated the accuracy of the HDNNP for bulk $\gamma$-belite.
For this purpose, the bulk structure was optimized by relaxing both, unit cell parameters and atomic positions, using the HDNNP as well as DFT. The obtained results are compiled in Table~\ref{table:bulk_parameters} together with available theoretical and experimental data from the literature. Overall, we find excellent agreement between the HDNNP and the DFT-optimized structures with lattice parameters showing relative errors of about 0.4~\% and only small deviations from experiment, which are within the typical errors of generalized gradient approximation functionals. The cohesive energy obtained with the HDNNP is -5.849~eV/atom, which is close to the DFT value of -5.851~eV/atom. Since the HDNNP has not been trained to free atom energies, the DFT energy of an isolated atom in its electronic ground state has been used in both cases to compute the cohesive energy.
Additionally, we performed an $NPT$ MD simulation for a ($3\times 2\times 3$) supercell to obtain the density of $\gamma$-belite at 300~K, which yields an equilibrium value of 2.82~g/cm$^3$, close to the experimental finding of 2.96~g/cm$^3$~\cite{P6886}. 

\begin{table}[htbp]
\centering
\caption{Bulk lattice parameters for $\gamma$-belite obtained from DFT and the HDNNP in this work. For reference, also experimental and DFT data (obtained with the PBE functional) from the literature is given.}
\label{table:bulk_parameters}
\begin{tabular}{c r r r r} 
\toprule		
 & \multicolumn{2}{c}{This work} & \multicolumn{2}{c}{Literature}\\
 & DFT & HDNNP & Exp.~\cite{P7158} & DFT~\cite{P6590} \\ \midrule
 \textit{a} ({\angstrom}): & 5.127 & 5.120 &	5.074 & 5.21 \\ 
 \textit{b} ({\angstrom}): & 11.422 & 11.312 & 11.211 & 11.37 \\
 \textit{c} ({\angstrom}): & 6.876 & 6.814 & 6.753 & 6.83 \\
\bottomrule	
\end{tabular}
\end{table}

\subsection{Simulations of $\gamma$-belite-water interfaces}

\subsubsection{Initial equilibration of the interfaces}

A key process for the equilibration of oxide-water interfaces is the saturation of undercoordinated interface atoms, which next to molecular adsorption also involves the dissociative adsorption of interfacial water molecules~\cite{P5601,P7208}. Specifically, there are two important processes at $\gamma$-belite-water interfaces, which are the adsorption of water molecules or hydroxide ions via the oxygen atoms at surface Ca atoms and the protonation of accessible oxygen atoms of SiO$_4$ groups. We have investigated these fast events during the initial equilibration MD simulations in the $NPT$ ensemble for both surface terminations.

With 16 Ca atoms in the first layer of the (4$\times$4) slab geometries, the T2 and T3 terminations each contain in total 32 under-coordinated Ca atoms, taking both slab surfaces into account. Based on the Ca under-coordinations in the topmost layer given in Table~\ref{table:cleavage_energy}, a full coordination of all Ca atoms by six oxygen atoms, like in the bulk material, requires the adsorption of an overall 96 oxygen atom-containing water species. 
Additionally, in the case of the T2 termination, the second Ca layer is not fully coordinated, as one oxygen neighbor is missing per Ca atom. However, in a bulk-like environment this oxygen atom is shared with the first-layer Ca atoms such that no additional oxygen species need to be adsorbed to reach a full coordination of the second Ca layer.
Moreover, both terminations contain under-coordinated oxygen atoms in overall 32 SiO$_4$ tetrahedra. Due to geometric constraints resulting from the specific rotation of the tetrahedra, the accessibility of these oxygen atoms strongly depends on the specific structure. For instance, at the ideal T2 termination one oxygen atom can be easily protonated.

Analyzing the time evolution of the adsorbed species observed in the equilibration trajectories (cf. Figs. S6 and S7 in the SI), we find that the average number of oxygen species adsorbed at the surface (Fig. S6a) is slightly lower for T3 (approximately 94) than for T2, which even exceeds 100. A reason for the overall larger number of adsorbed species at T2 is the more open surface structure showing trenches that provide easy access to the Ca atoms at the surface.

A further differentiation of the adsorbed oxygen atoms in molecular water and hydroxide ions resulting from water dissociation is presented in Fig. S7 in the SI, showing that for T2 about 30~\% of the adsorbed species are hydroxide ions, while for T3 the corresponding fraction is only about 8~\%. Since the total number of adsorbed water molecules is similar for both terminations, the main difference in the overall coordination of surface Ca atoms is due to the smaller amount of hydroxide ions present at T3 (cf. Fig. S6a in the SI). Comparing Figs. S6b and S7a in the SI, a strong correlation between the number of adsorbed hydroxide ions and protonated surface oxygen atoms can be observed. Thus, a large fraction of both water dissociation products is adsorbed at the surface. This dissociation of water molecules happens already in the first few picoseconds of the equilibration simulations for both terminations. Due to the very exposed geometry of the unsaturated oxygen atom on top of the SiO$_4$ tetrahedra at the T2 surface (cf. Fig.~\ref{fig:surfaces}) essentially all SiO$_4$ tetrahedra of this termination are protonated, which is an important reason for the increased water dissociation degree at this surface. Moreover, this increased level of water dissociation at T2 is also supported by a preference for hydroxide adsorption in the trenches close to the second Ca layer as discussed in more detail below. After 1~ns simulation time, the interfaces are fully equilibrated.

\subsubsection{Characterization of the hydrated surface \label{sec:NVT_results}}  

\textbf{Surface adsorption}

During the initial equilibration of the interface, different chemical species have been formed through water dissociation and proton transfer reactions. The time-averaged numbers of these species in the equilibrated systems obtained in 1~ns $NVT$ simulations are compiled in Table~\ref{table:number_of_chemical_species} for both terminations. Since hydroxide adsorption is only possible at the 16 calcium sites at each side of the slabs, on average 96\% of the Ca atoms are connected to a hydroxide ion for T2 while the respective value for T3 is only 24\%. Free hydronium ions in solution have not been found during the simulations. Therefore, the small difference between O$_\text{s}$H$^-$ and O*H$^-$ in Table~\ref{table:number_of_chemical_species} can be ascribed to non-absorbed hydroxide ions in solution that amount to about 6\% and 8\% of all OH$^-$ for T2 and T3, respectively. The vast majority of hydroxide ions formed by water splitting is thus adsorbed at the surface. 

The quantitative determination of the relative protonation degree of the surfaces is more challenging since the oxygen atoms in the SiO$_4$ groups show different accessibilities caused by different rotations of the tetrahedra. For instance, there is one protonated surface oxygen atom per tetrahedron at the T2 termination (32.5 protons at 32 SiO$_4$ groups, cf. Table~\ref{table:number_of_chemical_species}). As we will discuss below, these are the oxygen atoms pointing away from the surface, while the oxygen atoms at T3, which are coordinated by a silicon atom and two Ca atoms, show a much reduced affinity for protonation with only about 25~\% of the SiO$_4$ having a proton attached.

\begin{table}[htbp]
\centering
\caption{Time-averaged number of selected chemical species at both surfaces for the (4$\times$4) cells of the T2 and T3 terminations determined in MD simulations in the $NVT$ ensemble at 300~K. O$_\text{s}$H$^-$ refers to protonated surface oxygen atoms in SiO$_4$ tetrahedra, H$_2$O* to adsorbed water molecules, and O*H$^-$ to adsorbed hydroxide ions.}
\label{table:number_of_chemical_species} 
\begin{tabular}{l c c} 
\toprule		
 \multirow{2}{*}{Species:} & \multicolumn{2}{c}{Termination:}\\ \cline{2-3}
  & T2 & T3  \\ \midrule
 O$_\text{s}$H$^-$ & 32.5 & 8.2 \\
 H$_2$O*& 71.9 & 86.4  \\
 O*H$^-$& 30.7 & 7.6  \\
\bottomrule	
\end{tabular}
\end{table}

\textbf{Vertical density profiles}

\begin{figure*}
\centering
\includegraphics[width=1.0\linewidth]{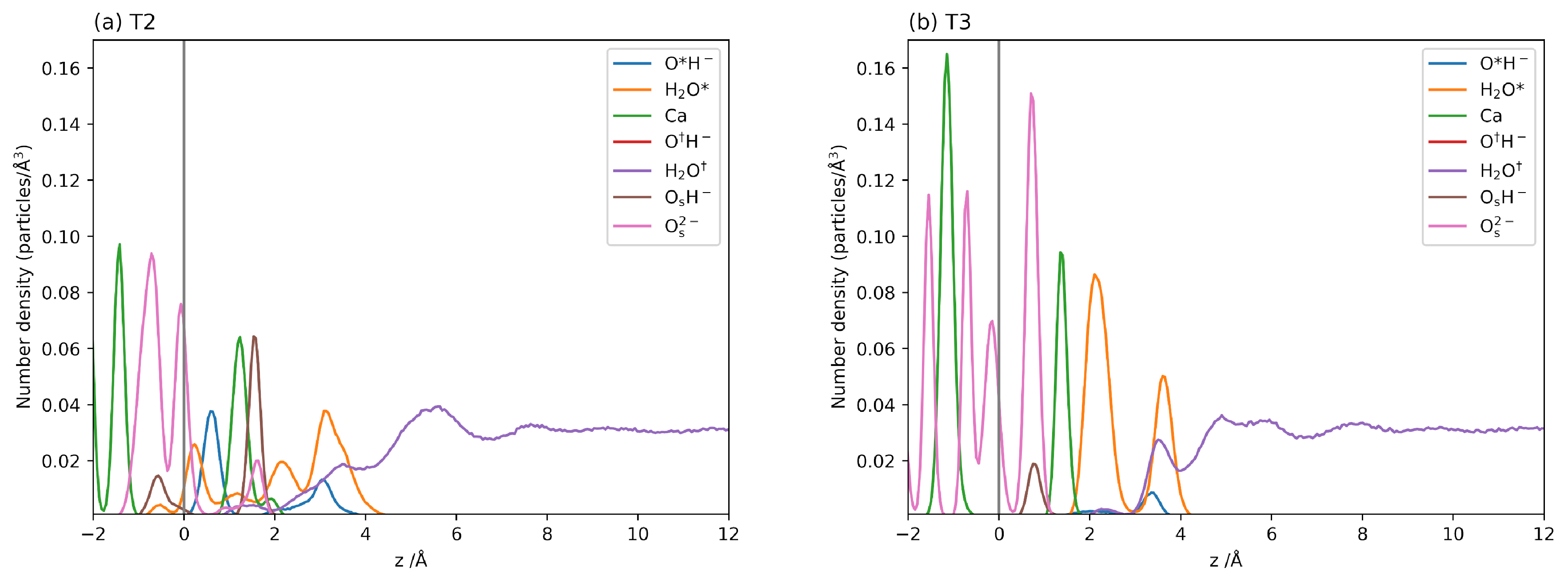}
\caption{Density profiles of different oxygen species (O*H$^-$ - adsorbed hydroxide ions, H$_2$O* - adsorbed water molecules, OH$^{-\dagger}$ non-adsorbed hydroxide ions, H$_2$O$^{\dagger}$ non-adsorbed water molecules, O$_\text{s}$H$^-$ - protonated surface oxygen atoms, O$_\text{s}^{2-}$ - non-protonated surface oxygen atoms) and calcium atoms along the $z$ ([010]) direction for the T2 and T3 terminations. $z=0$, indicated by a vertical grey line, corresponds to the average position of the outermost layer of silicon atoms in the MD simulations.}
\label{fig:density_profiles}
\end{figure*}

Figure~\ref{fig:density_profiles} shows the density profiles of different chemical species along the [010] direction perpendicular to the surface for both terminations relative to the position of the topmost Si atom layer defining $z=0$. The T3 termination (Fig.~\ref{fig:density_profiles}b) shows a clear outermost calcium peak at around 1.5~\AA{}, indicating the absence of any calcium dissolution. Closer to the surface, at about 0.75~\AA{}, a peak of protonated surface oxygen atoms can be found. Moreover, three hydration layers can be distinguished. The first layer (L1) up to 3~\AA{} contains mainly adsorbed water molecules (H$_2$O*) and tiny traces of oxygen species classified as adsorbed hydroxide ions and free water molecules. 
Apart from a relatively large amount of H$_2$O*, the second layer L2 in a range of 3-4~\AA{} consists of non-adsorbed water molecules and contains most of the adsorbed hydroxide ions in the vicinity of the top layer Ca atoms, which are exposed at the surface. Further, smaller peaks in the range of 4-9~\AA{} from the surface are less structured and consist of non-adsorbed water molecules only. Beyond a distance of about 9~\AA{}, the water molecules exhibit the constant density of bulk-like water. 

The density profile of the T2 surface (Fig.~\ref{fig:density_profiles}a) is characterized by a much larger number of overlapping peaks. Three peaks of adsorbed water molecules at a $z=-0.5$, $0.25$, and $1.2$~\AA{} are at distances smaller than or similar to the outermost main calcium peak at about 1.25~\AA{} and suggest an easy penetration of water molecules into the trenches at the surface. As discussed before, the more open surface geometry of T2 and under-coordinated calcium atoms of the second layer contribute to this by providing possible adsorption sites. 
Also, the majority of adsorbed hydroxide ions with a peak at about 0.6~\AA{} can be found here.
Two more peaks of adsorbed water molecules can be found at 2.15 and 3.15~\AA{} extending to about 4.5~\AA{}. Also in this region another albeit smaller peak of adsorbed hydroxide is located at 3.1~\AA{}. Moreover, for O$_\text{s}$H$^-$, two peaks are observed. Next to a dominant peak at about 1.5~\AA{} corresponding to the protonated outermost oxygen atoms of the SiO$_4$ tetrahedra, also a smaller peak of protonated surface oxygen atoms at -0.6~\AA{} deep in the surface is found. Small amounts of non-adsorbed water molecules can be found starting at 1~\AA{} from the surface culminating in a large peak at between 5 and 6 \AA{}. As for T3, water adopts a bulk-like density starting around 9~\AA{} from the surface. Interestingly, a small Ca peak slightly below 2~\AA{} indicates calcium displacement away from the topmost layer, which will be analyzed below.

\textbf{Lateral distribution of adsorbed species}

\begin{figure*}
\centering
\includegraphics[width=1.0\linewidth]{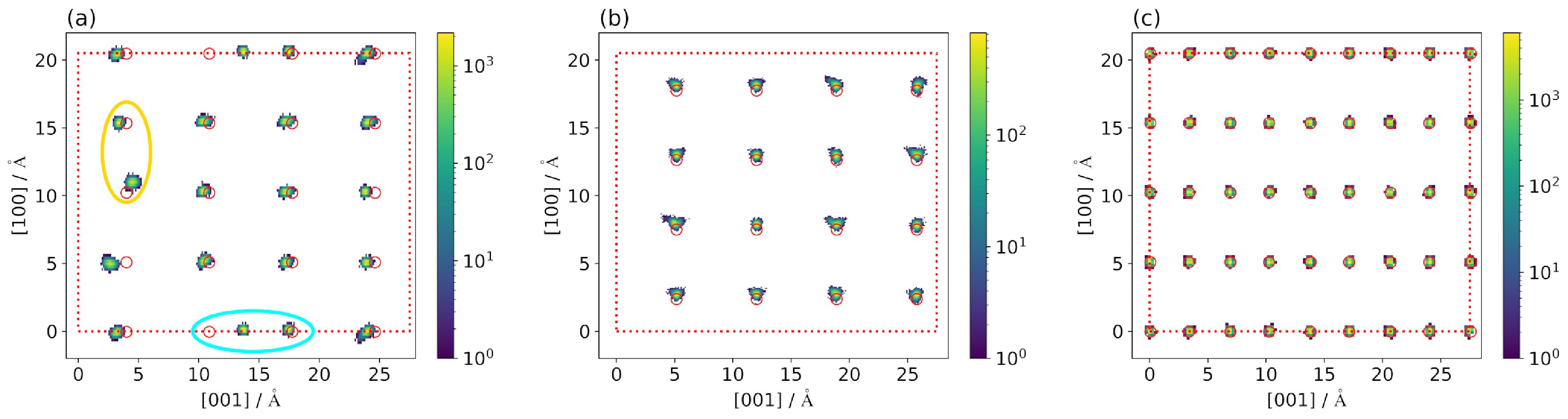}
\caption{Top views of the relative probability distributions of the calcium atom positions during $NVT$ MD trajectories at 300~K. The red circles represent the atomic positions at surfaces optimized in vacuum, the (4$\times$4) simulation cell is shown as a red dotted box. Panel (a) shows the first layer calcium atom distribution for T2, (b) the first layer of calcium atoms for T3, and (c) the second layer of calcium atoms for T3. The yellow and blue ellipses highlight two  Ca atom displacements at the T2 termination.}
\label{fig:Ca_positions}
\end{figure*}

In addition to the vertical density profiles, the lateral probability distributions of the different chemical species were analyzed. During the $NVT$ simulations, all Ca atoms (Fig.~\ref{fig:Ca_positions}) oscillate around the positions adopted during the initial $NPT$ equilibration, but interestingly in these equilibration runs some local surface changes could be observed. For instance, for the T2 termination, two Ca atoms showed notable displacements from their initial vacuum positions (cf. Fig.~\ref{fig:Ca_positions}a). The lower Ca atom in the yellow ellipse is shifted by about 1~\AA{} in the [101] direction forming a surface dimer with another Ca atom, which is accompanied by a rotation of a neighboring SiO$_4$ tetrahedron, we will from now on refer to as the ``type I'' displacement. 
Moreover, the left Ca atom in the blue ellipse is displaced along [001] by about 3.5~\AA{}, significantly reducing the distance to a neighboring calcium atom with an average final separation of only about 3.75~\AA{}. We will refer to this displacement as ``type II''.
Therefore, two different types of local structural changes on the surface could be observed for the T2 termination. 
In case of the T3 termination (Fig.~\ref{fig:Ca_positions}b), a small systematic shift of the topmost calcium layer in the [100] direction has been observed during the MD simulation. However, a drift of the entire slab can be excluded as the Ca positions in the second so not show any notable displacement (Fig.~\ref{fig:Ca_positions}c). This shift of the topmost Ca layer can be explained by the relaxation of the system due to the full coordination of the top layer calcium atoms by water species, which is reverting a corresponding shift in the opposite direction observed before when relaxing the bulk-truncated surface in vacuum.

\begin{figure*}
\centering
\includegraphics[width=1.0\linewidth]{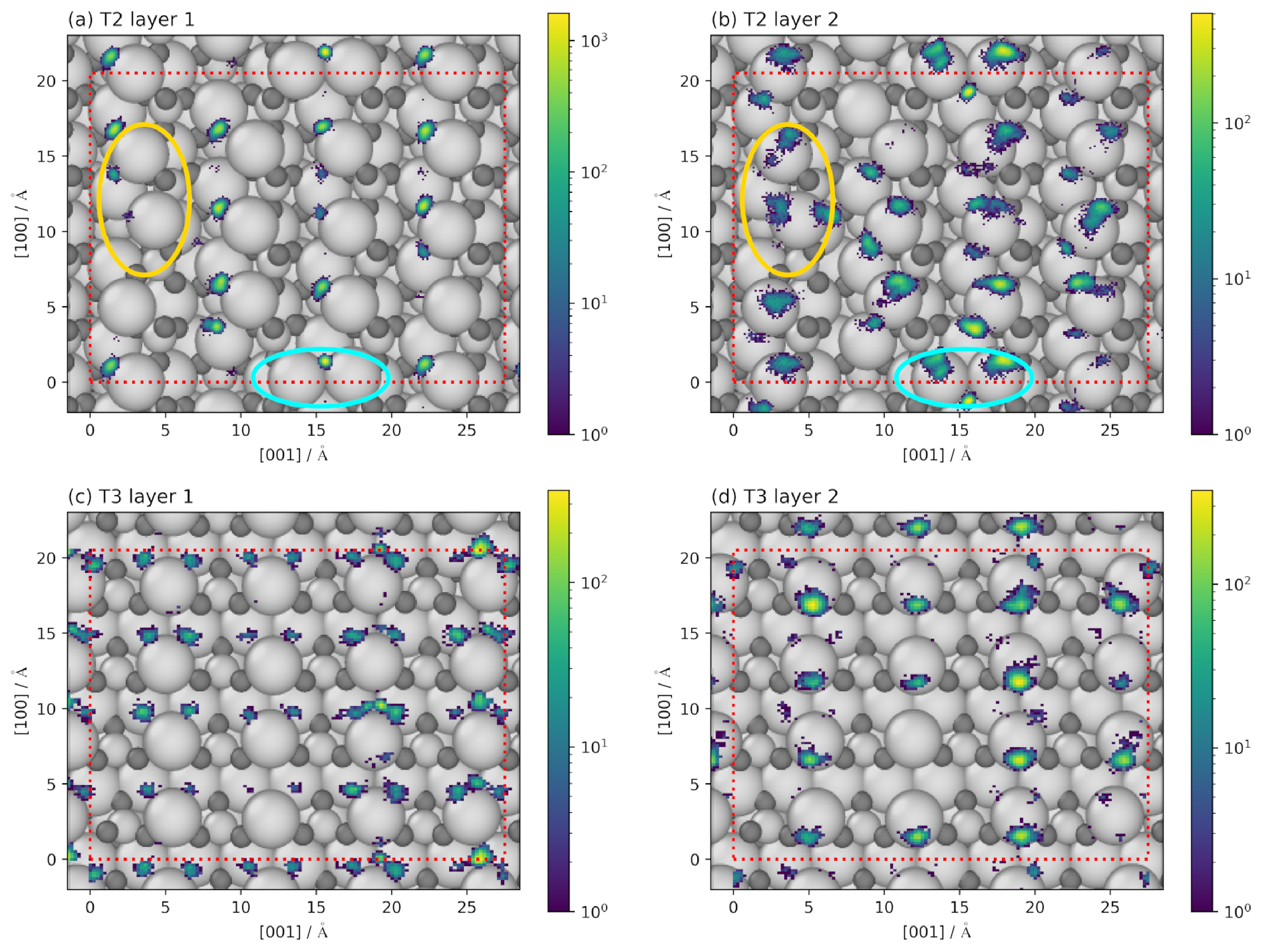}
\caption{Lateral probability distributions of the oxygen atoms in the adsorbed hydroxide ions for the T2 (a,b) and T3 (c,d) terminations. Panels (a) and (c) show the first and panels (b) and (d) the second layers of oxygen atoms (cf. blue peaks in Fig.~\ref{fig:density_profiles}). The positions of the surface atoms at the start of the $NVT$ simulations at 300~K are shown in grey, and the (4$\times$4) supercells used in the simulations are shown as red dotted boxes. The yellow and blue ellipses highlight two Ca atom displacements of type I and type II, respectively, observed for T2.}
\label{fig:Adsorbed_OH}
\end{figure*}

For both terminations, two layers of adsorbed hydroxide ions O*H$^-$ have been observed (cf. blue peaks in Fig.~\ref{fig:density_profiles}). The lateral distributions of the corresponding oxygen atoms are presented in Fig.~\ref{fig:Adsorbed_OH}. For T2 one preferred oxygen site is located in the first oxygen layer (Fig.~\ref{fig:Adsorbed_OH}a, blue peak at 0.6~\AA{} in Fig.~\ref{fig:density_profiles}a) with oxygen atoms bridging two Ca atoms in the first and second Ca layers, respectively. These atoms complete the six-fold coordination of the second-layer Ca atoms, and this site accommodates the majority of hydroxide ions in the system due to the electrostatically favorable position in between two Ca ions. Fig.~\ref{fig:Adsorbed_OH}b shows the distribution of the second O*H$^-$ layer on top of the surface of T2 (peak at 3.1~\AA{} in Fig.~\ref{fig:density_profiles}a), which is less well structured. For T3 two very weakly occupied adsorption positions on top of second-layer Ca atoms, which are already fully coordinated, can be identified for the first hydroxide layer (Fig.~\ref{fig:Adsorbed_OH}c), they correspond to a very small hydroxide peak at about 2~\AA{} in Fig.~\ref{fig:density_profiles}b. The distance of these hydroxide ions to the second Ca layer underneath is greater than 3~\AA{}, therefore, there is at most a very weak interaction. Instead, these hydroxide ions  are situated only slightly higher than the first Ca layer, but their lateral positions ensure the appropriate interaction distance to the first layer Ca atoms. Due to the large separation between the Ca atoms in the first layer, which does not allow the oxygen atoms to adopt bridging positions in Ca dimers within the first layer, the binding of these hydroxide ions is very weak, such that the first-layer Ca atoms cannot adopt a static six-fold oxygen coordination through adsorption of only O*H-.
A well-structured contribution to the coordination of the first layer Ca atoms at T3 is achieved by strongly ordered adsorbed hydroxide ions on top of the first layer Ca atoms (Fig.~\ref{fig:Adsorbed_OH}d, cf. blue peak at 3.35~\AA{} in Fig.~\ref{fig:density_profiles}b). Two first layer Ca atoms in Fig.~\ref{fig:Adsorbed_OH}d do not show attached hydroxide ions in the second layer.

\begin{figure*}
    \centering
\includegraphics[width=1.0\linewidth]{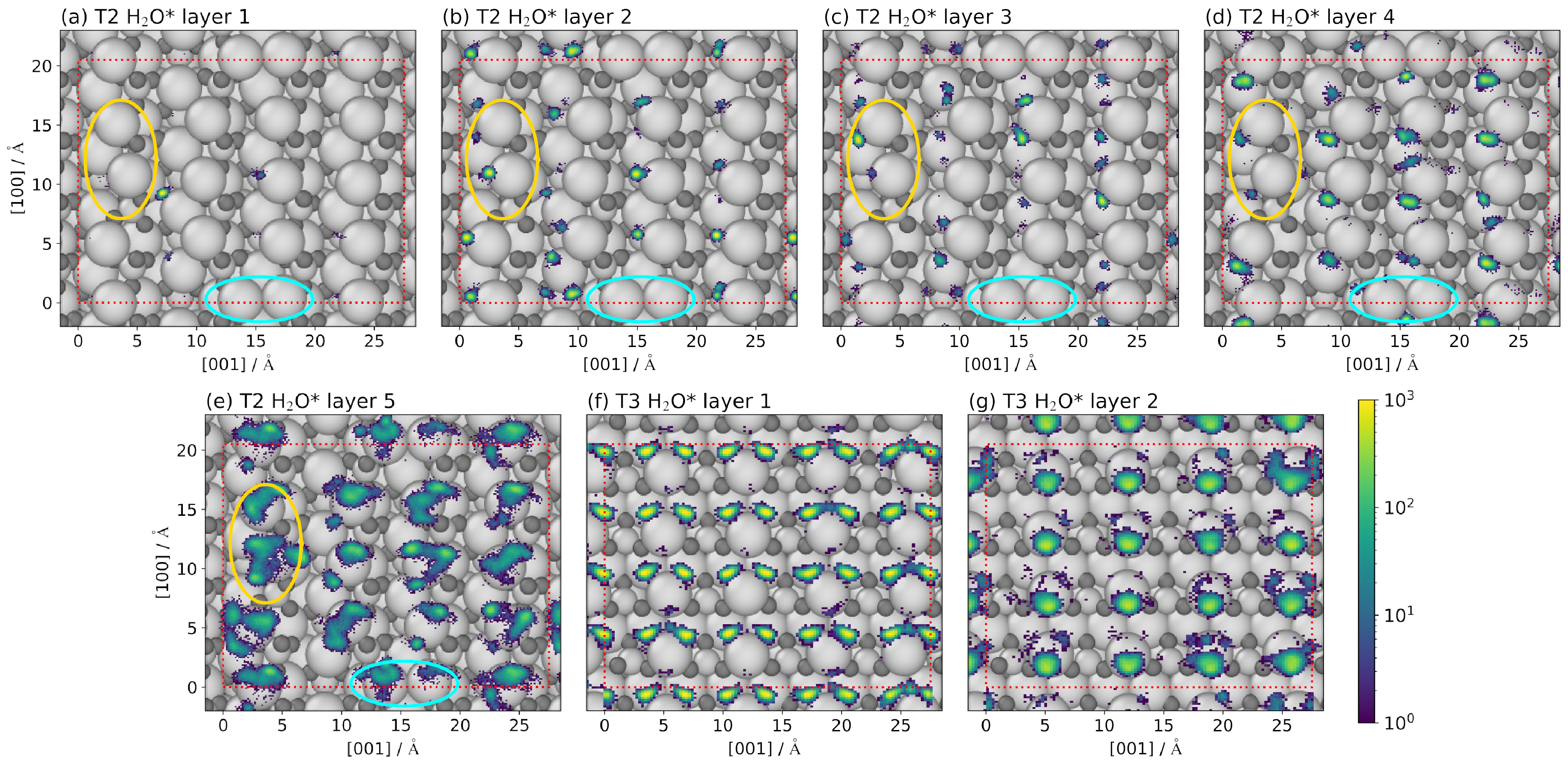}
    \caption{Lateral probability distributions of oxygen atoms in adsorbed water molecules (with Ca-O distances smaller than 2.9~\AA{}) for the T2 (panels a-e) and T3 (panels f and g) terminated surfaces. The grey atoms represent the starting positions of the surface atoms in the \textit{NVT} MD simulations. 
    The (4$\times$4) simulation cells are shown as a red dotted boxes. The yellow and blue ellipses highlight two Ca atom displacements of type I and type II, respectively, observed for T2.}
    \label{fig:T2_T3_H20}
\end{figure*}

For the vertical density profiles of molecular water oxygen atoms at T2 (orange curves in Fig.~\ref{fig:density_profiles}a), five peaks have been observed. Two of these are situated below the first layer of calcium atoms, one is at about the same height, and two are located above the surface. The lateral distributions of the oxygen atoms corresponding to these peaks are shown in Figure~\ref{fig:T2_T3_H20}. 
The first peak, deepest inside the surface, can be identified as a single isolated oxygen adsorption site at a second layer Ca atom that is formed by the rotation of a SiO$_4$ tetrahedron in a type I defect (Fig.~\ref{fig:T2_T3_H20}a). The lateral distribution of the water molecules corresponding to the second peak shown in Fig.~\ref{fig:T2_T3_H20}b shows a bridging coordination of two Ca atoms in the first and second layer, respectively, similar to the hydroxide ions in Fig.~\ref{fig:Adsorbed_OH}a, although the water molecules exhibit overall a less well ordered structure. Also, the relative peak heights in Fig.~\ref{fig:density_profiles}a show that the contribution of hydroxide ions is dominant over water molecules. The third water layer at about the same height as the first Ca layer in Fig.~\ref{fig:T2_T3_H20}c shows very little structure and contains only a small number of molecules (cf.~Fig.~\ref{fig:density_profiles}a). The on-surface water positions in layers 4 and 5 are, however, well structured (Figs.~\ref{fig:T2_T3_H20}d,e) to complete the six-fold oxygen coordination of the first-layer Ca atoms. The two adsorbed water layers at the T3 termination (Figs.~\ref{fig:T2_T3_H20}f,g) are very strongly structured with overall three oxygen positions to achieve a full coordination of the topmost Ca layer at this surface.

\begin{figure}
\includegraphics[width=1.0\linewidth]{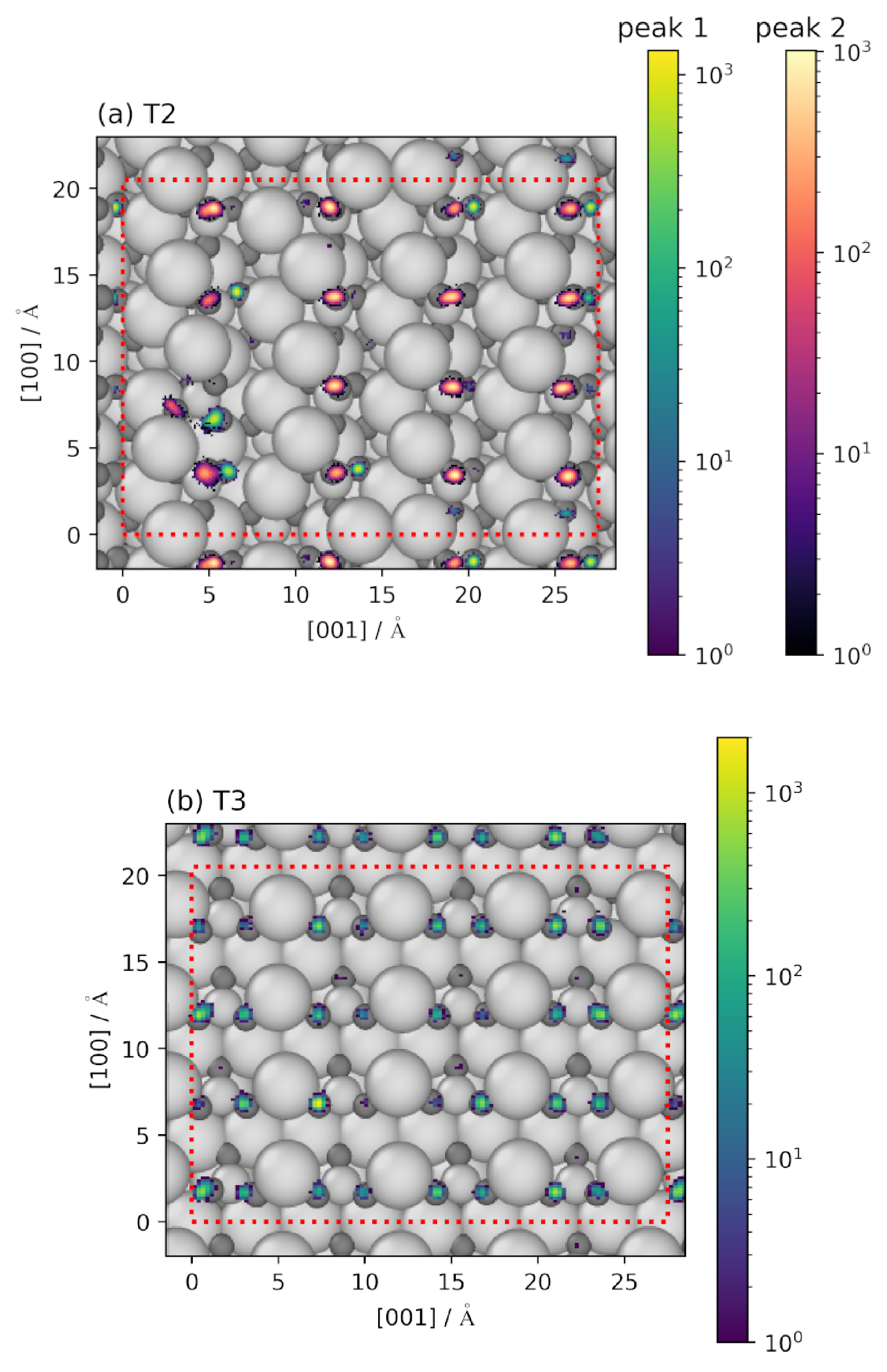}
\caption{Lateral distributions of the protonated surface oxygen atoms for the T2 (a) and T3 (b) terminations. The initial positions of the surface atoms for the \textit{NVT} MD simulation are shown in grey, the (4$\times$4) simulation cells are shown as red dotted boxes. For T2, the two different peak positions (cf. Fig.~\ref{fig:density_profiles}a) are represented by two different color scales.}
\label{fig:OsH}
\end{figure}

Finally, we have analysed the lateral position of the protonated surface oxygen atoms in Fig.~\ref{fig:OsH}. As can be seen in Fig.~\ref{fig:density_profiles}, there are two peaks for T2 and one peak for T3. For T2 (Fig.~\ref{fig:OsH}a) the first peak is not present for all sites and corresponds to the protonation of a less accessible oxygen atom deeper in the surface. The main second peak for the protonated oxygen atoms on top of the SiO$_4$ tetrahedra is strongly pronounced for all sites with the only exception of a rotated tetrahedron due to the type I surface displacement of a Ca atom. At the T3 surface (Fig.~\ref{fig:OsH}b), protonation is observed for the two topmost oxygen atoms of the SiO$_4$ tetrahedra, which have the same environment, while the third oxygen atom a bit further away from the surface is less accessible and thus rarely protonated. 

\textbf{Coordination of calcium atoms on the surface}

\begin{figure}
\centering
\includegraphics[width=1.0\linewidth]{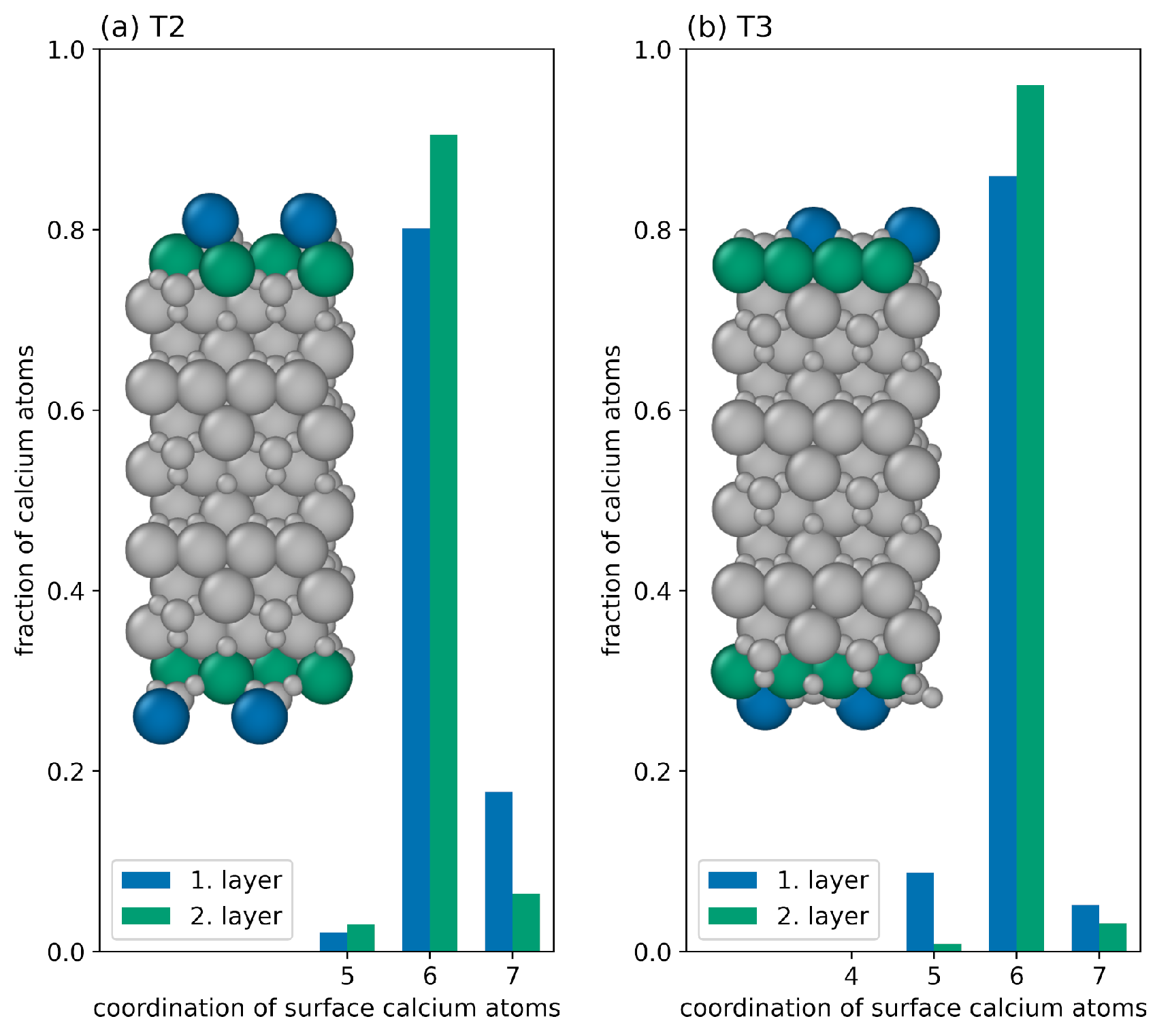}
\caption{Coordination number distributions of the calcium atoms by oxygen atoms, both from surface oxygen and water molecules, in the first two layers highlighted by the insets for the T2 (a) and T3 (b) terminations.}
\label{fig:Ca_coord_new}
\end{figure}

Calcium atoms in bulk $\gamma$-Belite are six-fold coordinated by oxygen atoms, whereas on the free surfaces the coordination is significantly lower (cf. Table~\ref{table:cleavage_energy}). The coordination number distributions of the calcium atoms in the first two layers considering oxygen atoms of all species up to a distance of 2.9~\AA{} at the water-$\gamma$-belite interfaces, are shown in Fig.~\ref{fig:Ca_coord_new}. For both terminations, the vast majority of Ca atoms is six-fold coordinated. In case of the T2 termination, almost 20\% for the first layer Ca atoms, which are very exposed to the aqueous phase, are formally even seven-fold coordinated. This high calcium coordination has also been observed in the polymorph $\beta$-belite~\cite{P6878} and is also known to be present to some extent in solutions of Ca ions in water~{\cite{P7165}}. A surprisingly high fraction of only five-fold coordinated Ca atoms is found for the first layer at T3, which can be explained by the very dynamic solvation structure of the water molecules around these atoms with a large in-plane Ca-Ca distance preventing the formation of a stable solvation structure.

\subsubsection{Surface defects and reconstruction of T2}

As discussed above, for the T2 termination two types of calcium atom displacements have been observed during the $NPT$ equilibration of the solid-liquid interfaces, which indicate the possible existence of more favorable structures and the onset of surface reconstructions, which might be induced and stabilized by the solvent. Before we will further investigate these structural changes, we note that after the initial $NPT$ equilibration the surface structures were preserved in the $NVT$ trajectories and no further Ca displacements have been observed. Therefore it is likely that structural changes at the surface are connected to high barriers, such that more stable structures than those observed here at the surface may exist.

\textbf{Type I}

\begin{figure}
\centering
\includegraphics[width=0.99\linewidth]{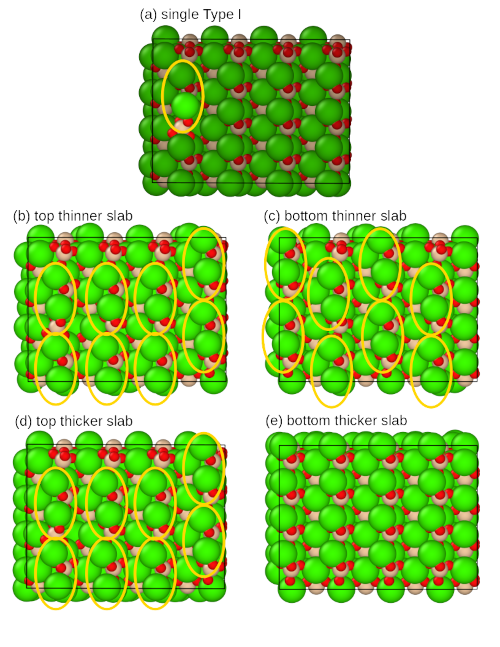}
\caption{(a) Top views of the T2 terminated surfaces after introducing a single Type I defect on one side of the slab. Displaced atoms are shown with lighter colors. Two different thicknesses of the (4$\times$4) supercells have been used, corresponding to a single bulk unit cell (b,c) and two bulk unit cells (d,e). Panels (b) and (d) show the top surfaces of the slabs, (c) and (e) the respective bottom surfaces after optimization. The atom pairs adopting a type I displacement are marked by yellow ellipses.}
\label{fig:TypeI}
\end{figure}

The first type of defect involves the displacement of a Ca atom and the rotation of an adjacent SiO$_4$ tetrahedron. To investigate the stability of this structural change of the surface in the absence of the solvent, we generated such a point defect of type I on one side of the slab based on the surface optimized in vacuum (Fig.~\ref{fig:TypeI}a). Starting from this surface, slabs of two different thicknesses, 12.6 and 24.1~\AA{} corresponding to one and two bulk unit cells, respectively, were optimized by DFT in vacuum. The top and bottom sides after the optimization are shown in Fig.~\ref{fig:TypeI}b-e for both slabs. In both cases, the single initial point defect initiated the formation of numerous further type I defects converting the top surfaces entirely (Figs.~\ref{fig:TypeI}b,d). Moreover, for the thin slab, the initial defect on the top surface also resulted in the formation of a complete layer of similar defects at the bottom surface (Fig.~\ref{fig:TypeI}c). A closer inspection of the formed structures shows that in principle several arrangements of the defects, e.g. a ``parallel''  and a ``zig-zag'' pattern, are possible and we leave a closer investigation for future work. In case of the thicker slab the bottom surface did not change and remained in the optimized geometry of the bulk-truncated slab (Fig.~\ref{fig:TypeI}e). This transport of information from one side of the thin slab to the other clearly shows that a sufficient slab thickness of at least two bulk unit cells is required to decouple both surfaces in simulations. 

Both slabs are energetically more stable in vacuum than the optimized bulk-truncated surfaces. For the thin slab, the cleavage energy is reduced from 1.49~J/m$^2$  to 1.45~J/m$^{2}$ corresponding to a total energy reduction of 0.669 meV per atom in the slab. For the thick slab, the average cleavage energy reduces from 1.50 to 1.49~J/m$^{2}$, but we note that this reduction is a consequence of structural changes on the top surface only, and consequently the effective stabilization of the transformed surface is higher. The deviations from the cleavage energy values reported in Table~\ref{table:cleavage_energy} are related to the reduced thicknesses of the slabs investigated here.

\textbf{Type II}

\begin{figure}
    \centering
\includegraphics[width=1.00\linewidth]{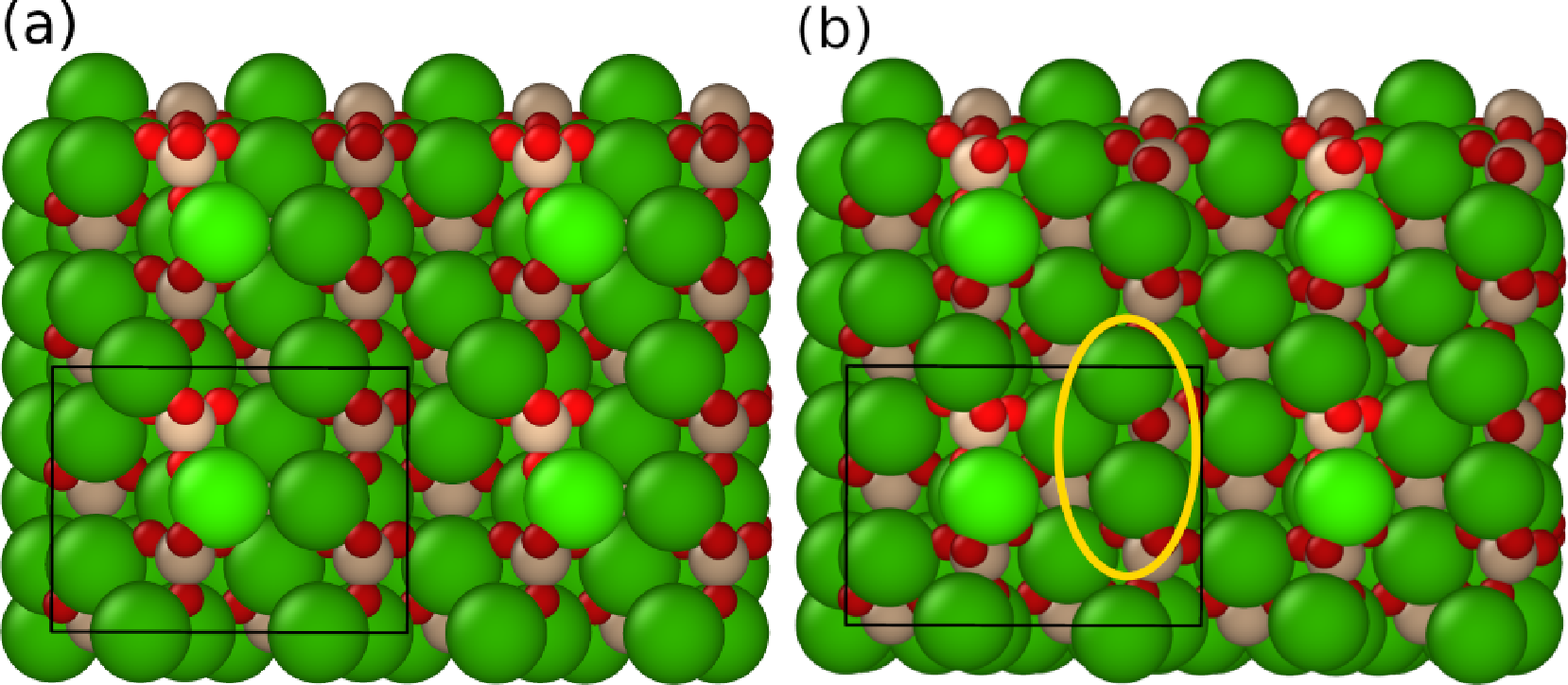}
    \caption{Top view of the T2 terminated surface in vacuum including a type II Ca dimer before (a) and after (b) a DFT geometry optimization. The optimized (2$\times$2) supercell of T2 is marked by the black rectangle, and to illustrate the structure several periodic images are shown. The displaced Ca atom is highlighted in light green and the yellow ellipse labels the geometry of a Ca pair adopting a geometry similar to a type I defect after optimization.}
    \label{fig:TypeII}
\end{figure}

The second defect observed on the T2 surface in the presence of water, type II, involves the displacement of a Ca atom towards a neighboring first-layer Ca atom position to form a Ca dimer with an interatomic distance of 3.75~\AA{} at the surface. Figure~\ref{fig:TypeII}a shows this type II defect inserted in the DFT-optimized bulk-truncated surface in vacuum. While the distance between two first-layer Ca atoms at the ideal T2 surface before forming the dimer is 6.88~\AA{}, and the overall closest Ca-Ca interaction between first-layer atoms is 5.13~\AA{}, the distance in the created calcium dimer is shorter and very similar to the Ca-Ca distances in the rows in the calcium-terminated T1 surface (cf. Fig.~\ref{fig:surfaces}), which are alternatingly 3.53 and 3.34~\AA{}. In principle, the displacement of the Ca atom in the type II defect found in the equilibration simulations in the presence of water could thus be interpreted as a Ca hopping event into an unoccupied site of an incomplete Ca row at the T2 surface. The result of the DFT-based geometry optimization of the type II defect in vacuum is shown in Fig.~\ref{fig:TypeII}b. As can be seen, in the absence of water the system adopts a geometry similar to a type I defect at the T2 surface, although all SiO$_4$ tetrahedra are distorted to some degree suggesting that a full transition to a type I-covered surface might be the most favorable surface structure, which is hindered by some barriers in the present case. Therefore, it can be concluded that the type II defect is unstable in the absence of water at the surface. 
Instead, the presence of strong O*H$^-$  and H$_2$O* peaks between the two Ca atoms of the dimer (Fig.~\ref{fig:Adsorbed_OH} and Fig.~\ref{fig:T2_T3_H20}) show that a stabilization of the type II defect by bridging oxygen atoms is essential. 

\begin{figure}
    \centering
    \includegraphics[width=0.35\textwidth]{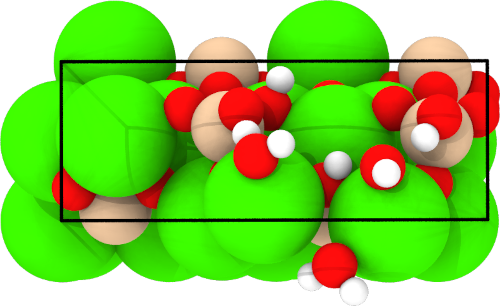}
    \caption{Top view of the DFT-optimized T2 terminated $(2\times 1)$ surface cell containing a type II defect in vacuum.
Four water molecules have been found necessary to stabilize the Ca dimer. Two of these molecules dissociate during adsorption while the other two remain intact.}
    \label{fig:TypeII+4H2O}
\end{figure}

To further explore this stabilization, we have performed DFT calculations to determine the minimum number of water molecules that are necessary to stabilize the Ca dimer of the T2 defect at the surface. As shown in Fig.~\ref{fig:TypeII+4H2O}, we found that four molecules are sufficient. Two of these molecules are dissociated with hydroxide ions adsorbing at the Ca atoms. The combination of four oxygen atoms originating from the water molecules and six oxygen atoms of the surface result in overall ten coordinating oxygen atoms, which match the number required to fully coordinate the Ca atoms in the dimer corresponding to two edge-sharing octahedra. This surface structure is further stabilized by the two protons adsorbing at the SiO$_4$ tetrahedra resulting in a full protonation of all available oxygen sites at the surface. 

\begin{figure}
    \centering
    \includegraphics[width=1.00\linewidth]{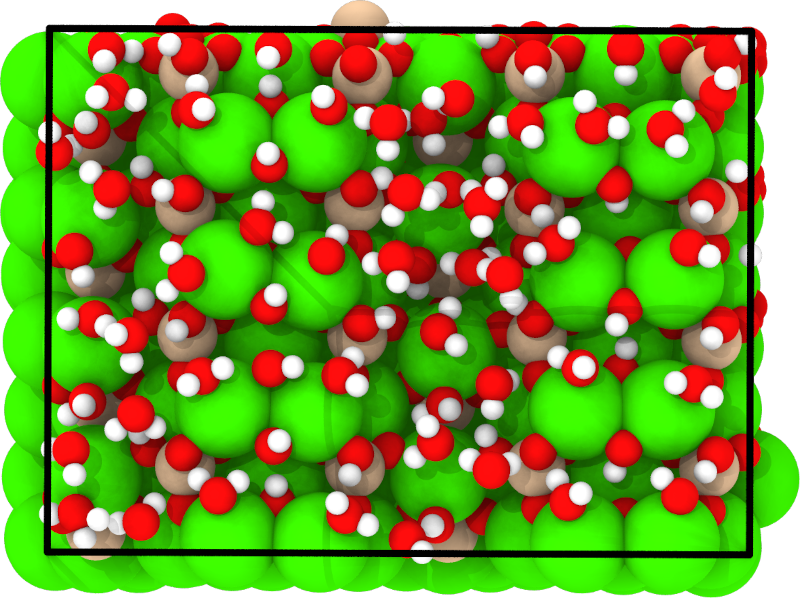}
    \caption{Top view of a (4$\times$4) supercell of the T2 terminated surface completely covered by Ca dimers of the type II reconstruction. Shown is the structure obtained after a 100~ps MD simulation at 300~K employing the HDNNP in the \textit{NPT} ensemble for the surface in contact with bulk water.}
    \label{fig:TypeII_HDNNP}
\end{figure}

Like in case of the type I defect, it is interesting to investigate if the formation of the dimer can be interpreted as the onset of a global transition of the surface. To study the stability of a fully converted surface we have displaced every second Ca atom of the first layer of a (4$\times$4) supercell such that the surface is completely covered by dimers in a rectangular pattern. The surface has then been combined with a liquid water phase consisting of 576 molecules and HDNNP-based MD simulations have been performed for the solid-liquid interface in the $NPT$ ensemble. To avoid a decomposition of the surface during the equilibration of the system, the surface atoms have initially been frozen for 10~ps to first relax the water structure at the interface. Then, a  simulation has been run for 100~ps with fully mobile atoms. The final structure of the simulation displayed in Fig.~\ref{fig:TypeII_HDNNP} shows that all dimers remain stable under simulation conditions. Two bridging hydroxide ions and/or water molecules are observed for every Ca atom dimer along with another molecule at each Ca atom resulting in a similar solvation pattern as in the vacuum optimization including four water molecules. Also different patterns of dimer structures like hexagonal arrangements are in principle possible and should be investigated in future work.

\textbf{Full/empty-row reconstruction}

\begin{figure}
\includegraphics[width=0.5\linewidth]{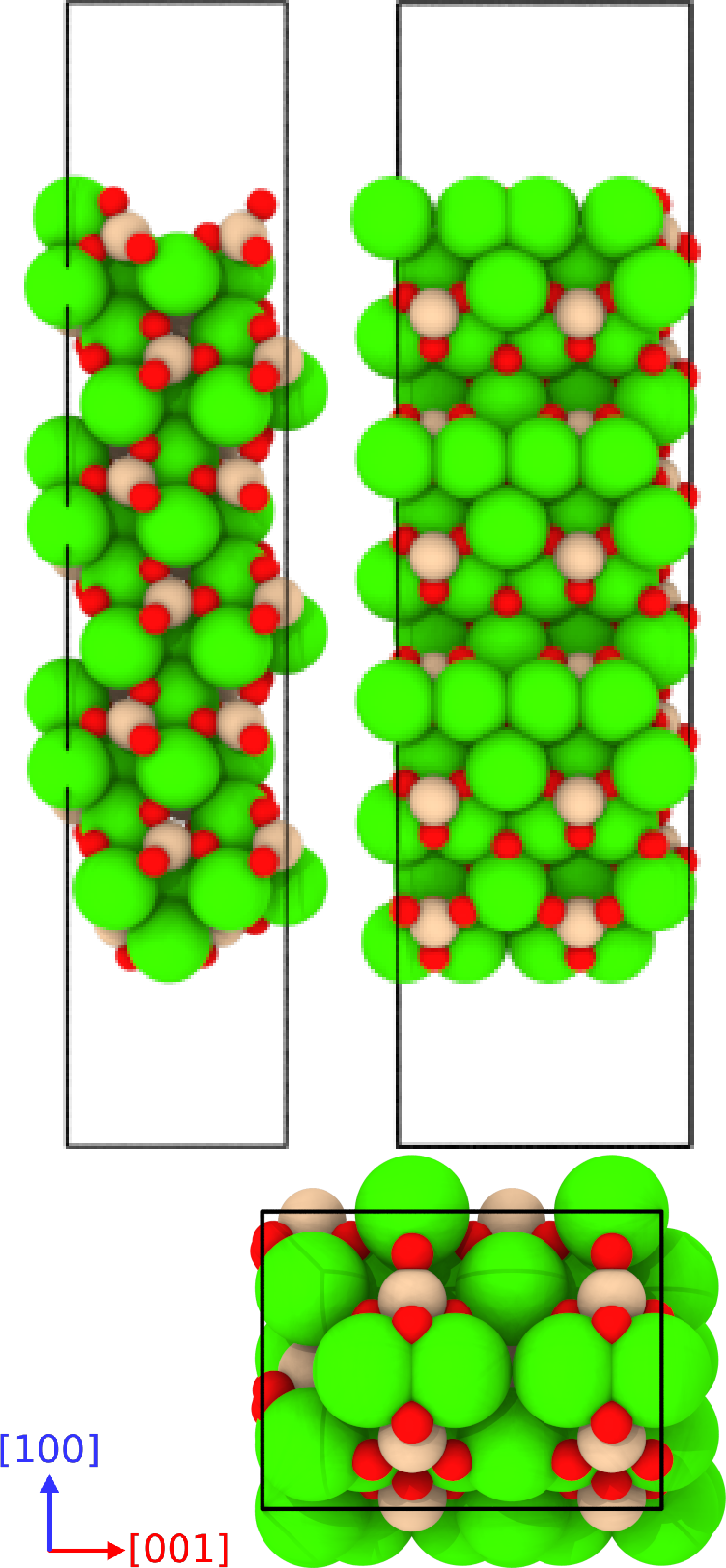}
    \caption{(2$\times$1) supercell of the full/empty-row termination T5 (corresponding to a ($4\times 2$) supercell of the bulk-truncated T2 termination). The first row presents side views along the [100] and [001] directions. Both surfaces are equivalent, and the last row shows a bottom view. }
    \label{fig:T5}
\end{figure}

\begin{figure}
    \centering
    \includegraphics[width=0.48\textwidth]{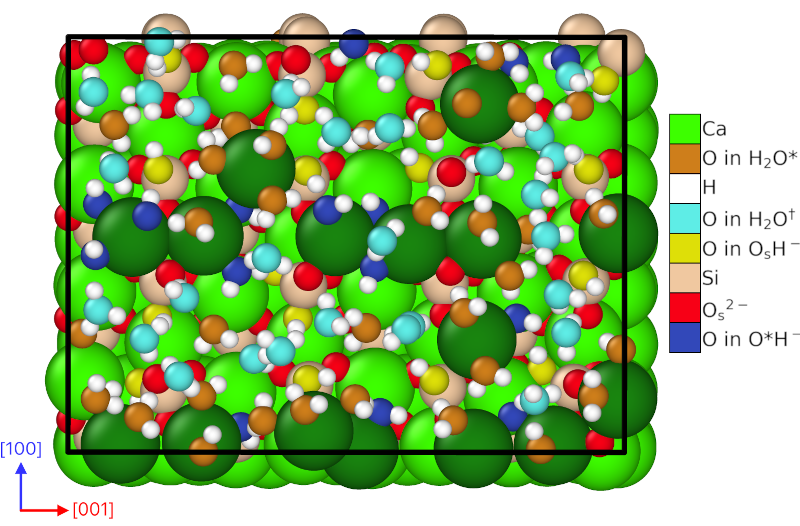}
    \caption{Top view of the T5-water interface after a HDNNP-driven 100~ps MD simulation in the \textit{NPT} ensemble. The Ca atoms of the first layer are shown in dark green. The simulation box was built as a (4$\times$4) supercell of the single unit, which correspond to (4$\times$2) supercell of the slab shown in Fig.~\ref{fig:T5}.}
    \label{fig:T5_with_H2O}
\end{figure}

As the formation of Ca dimers at the surface seems to be energetically favorable, we have further explored if a rearrangement of all first layer Ca atoms in alternating fully occupied and completely empty rows along the [001] directions is stable. Similar but denser rows are present at the T1 termination, which, however, is much less stable due to the polar calcium-terminated and oxygen-terminated surfaces. The structure of the investigated T5 termination is shown in Fig.~\ref{fig:T5}. This nonpolar surface structure can be derived either from T1 by transferring every second row of Ca atoms from the top to the bottom surface (cf. Fig.~\ref{fig:surfaces}a) or by redistributing the atoms of the dimers in the type II reconstruction of the T2 termination to completely filled and empty rows at both surfaces. 

First, we have computed T5 by DFT in vacuum resulting in a cleavage energy of 1.87~J/m$^2$, which is only slightly larger than the cleavage energy of T2 (cf. Table~\ref{table:cleavage_energy}) but much more stable than T1 and T4. This only slightly lower stability in vacuum compared to T2 and T3 is related to the similar coordination of the surface atoms in case of T5.
Still, as is evident in Fig.~\ref{fig:T5}, there is a lack of bridging oxygen atoms in the Ca rows that would be required to further stabilize T5. Therefore, as a next step we have tested the stability of T5 by HDNNP-driven MD simulations in the $NPT$ ensemble in liquid water to complete the coordination of the Ca atoms corresponding to long chains of edge-sharing octahedra. Although not trained to T5-like structures, we found the HDNNP to be transferable to simulations of the T5 solid-liquid interface enabling the calculation of long MD trajectories. Still, for more detailed studies of this surface a refinement of the potential should be considered. The structure observed at the end of a 100~ps MD simulation in Fig.~\ref{fig:T5_with_H2O} shows that the favorable six-fold coordination of the Ca atoms by oxygen atoms in water molecules and hydroxide ions is formed as expected. On the other hand, the Ca rows do not remain intact indicating that the full/empty row structures is not maintained in an aqueous environment. 

\section{Conclusions}\label{sec:conclusions}

In this work, we have investigated the aqueous interfaces of the two most stable non-polar T2 and T3 terminations of the $\gamma$-belite (010) surface using a high-dimensional neural network potential, which has been trained to density functional theory data. Employing this potential, the structure and reactivity of the interfacial water as well as structural changes of the surfaces have been studied by large-scale molecular dynamics simulations. 

For all interfaces we find that water can adsorb molecularly or dissociatively. The degree of dissociation is mainly governed by the need to saturate undercoordinated surface atoms with the exposed dangling bonds of oxygen atoms of the SiO$_4$ tetrahedra playing a dominant role. As a consequence, essentially all protons emerging from water dissociation are adsorbed at these oxygen atoms at the surface. The majority of the simultaneously formed hydroxide ions is attached to undercoordinated surface Ca atoms, whose coordination spheres are completed by additional water molecules. This finding explains the lower degree of water dissociation at T3 compared to T2, since the T2 surface exhibits easily accessible unsaturated oxygen atom in SiO$_4$ tetrahedra leading to increased dissociation at this surface.

In our simulations, the solid-liquid interface of the T3 termination, which is also most stable in vacuum, remains intact on nanosecond time scales. Contrarily, for the T2 termination in water two types of structural changes have been observed early in the $NPT$ equilibration simulations. 
The ``type I'' defect involves the displacement of a Ca atom in combination with the rotation of a SiO$_4$ group. This structure is also stable in vacuum and geometry optimizations in vacuum including a seed defect result in the conversion of the entire surface, i.e., a reconstruction with a lower cleavage energy than the original optimized bulk-truncated T2 termination. For thin slabs this reconstruction even extends to the opposite surface of the slab underlining the importance of using a sufficient slab thickness in simulations to decouple both surfaces.

The second reconstruction ``type II'' is the formation of a Ca surface dimer, which is only stable in water, since its fully coordinated form of two edge-sharing CaO$_6$ octahedra requires four additional water molecules and/or hydroxide ions, which are not available at the bulk-truncated surface in vacuum. When including this minimum number of four water molecules, a stabilization is also possible in vacuum, which goes along with a dissociation of two of these molecules simultaneously enabling the protonation of two unsaturated SiO$_4$ tetrahedra. In bulk liquid water, also a complete reconstruction of the entire surface covered by type II motifs remains stable in MD simulations but does not spontaneously form. 

For both reconstructed surfaces different geometric patterns of the type I and type II structural features are possible and are likely to exhibit similar energetic stabilities. 
The persistence of all observed surface features in long $NVT$ simulations indicates the presence of substantial barriers for surface structural changes, which have been formed only under the relatively harsh conditions of the initial $NPT$ equilibration simulations. All these findings suggest the existence of a considerable structural variety of surface motifs at $\gamma$-belite-water interfaces. As the stability of these structures is strongly influenced by the presence of water, revisiting terminations like T1 and T4, which are unfavorable in vacuum, and their possible reconstructions will be an interesting topic of future work.

\section{Acknowledgments}

This project has been funded by the Deutsche Forschungsgemeinschaft (DFG, German Research Foundation) under Germany’s Excellence Strategy EXC 2033 RESOLV (project-ID 390677874).
The authors gratefully acknowledge the computing time made available to them on the high-performance computer Noctua2 at the NHR Center PC$^2$. This center is jointly supported by the Federal Ministry of Education and Research and the state governments participating in the National High-Performance Computing (NHR) joint funding program (\url{http://www.nhr-verein.de/en/our-partners}).

\section{Conflict of Interest}

There are no conflicts of interest to declare.


\clearpage

\setstretch{1.0}
\onecolumngrid
\begin{center}
{\large\bfseries Supporting Information: Molecular Dynamics Simulations of $\gamma$-Belite(010)-Water Interfaces with High-Dimensional Neural Network Potentials} \\[1em]

Bernadeta Prus and J\"{o}rg Behler \\[0.5em]

\texttt{joerg.behler@rub.de} \\[0.5em]

\begin{itshape}
{\footnotesize
Lehrstuhl f\"ur Theoretische Chemie II, Ruhr-Universit\"at Bochum, 44780 Bochum, Germany \\
Atomistic Simulations, Research Center Chemical Sciences and Sustainability, Research Alliance Ruhr, 44780 Bochum, Germany
} \\[1em]
\end{itshape}
\today
\end{center}
\twocolumngrid

\begin{appendices}
\setcounter{figure}{0}
\renewcommand{\thefigure}{S\arabic{figure}}
\setcounter{table}{0}
\renewcommand{\thetable}{S\Roman{table}}

\section{Density Functional Theory Calculations}

The DFT reference calculations for training the high-dimensional neural network potentials (HDNNP) have been performed using the all-electron FHI-aims code (version 221103)~\cite{P2189}. The RPBE functional~\cite{P0120} has been chosen to describe electronic exchange and correlation in combination with light settings 
for the numerical atomic orbital basis set and integration grids. For the periodic systems, a \textbf{k}-point density of 4/{\angstrom}$^{-1}$ was chosen resulting in a \textbf{k}-point grid of 5$\times$3$\times$4 for the 28 atom bulk unit cell of $\gamma$-belite with dimensions of 5.134$\times$11.211$\times$6.753\AA{}$^3$~\cite{P7158}. For the slab calculations, the \textbf{k}-point grids have been adjusted to maintain the same density.

Before training the HDNNP, dispersion corrections have been added to the energies and forces employing the D3 method of Grimme using the DFT--D3 v.3.1 code~\cite{P3112} with zero damping, since RPBE-D3 has been demonstrated to provide a good description of liquid water~\cite{P3875,P4411,P4556,P7126} and also of bulk cementitious minerals~\cite{P7159}. 
The DFT-based structural optimizations were carried out through the Atomic Simulation Environment (ASE)~\cite{P5915} with a convergence criterion of 0.05 eV/{\angstrom}, no symmetry constraints were applied.

\section{Molecular Dynamics Simulations}

Molecular Dynamics (MD) simulations were carried out with the Large-scale Atomic/Molecular Massively Parallel Simulator (LAMMPS - version from 15 Jun 2023)~\cite{Thompson_2022} in combination with the Neural Network Potential Package (n2p2 v2.2.0)~\cite{Singraber_2019}, which allows to use energies, forces and the stress tensor provided by HDNNPs trained with \emph{RuNNer}~\cite{Behler_2015,Behler_2017}. Depending on the property of interest, the MD simulations were run in the $NPT$ or $NVT$ ensemble. A Nosé--Hoover thermostat (300~K) and/or barostat (1~bar) was applied, with the barostat coupled only to the surface normal direction for simulations of solid-liquid interfaces to maintain the bulk lattice constant in the interior of the $\gamma$-belite slabs. The MD time step was set to 1~fs for bulk $\gamma$-belite and to 0.5~fs for bulk water as well as the solid-liquid interface structures. In the case of solid-liquid interfaces, to prevent movement of the slab in the $x$ and $y$ directions, the total momentum of the central part of the belite slab with a diameter of 9~\AA{} was fixed in those directions.

MD simulations of bulk $\gamma$-belite were performed for (3$\times$2$\times$3) supercells containing 756 atoms in the $NPT$ ensemble for 1~ns. 
For MD simulations of the water/belite interfaces initial slabs were prepared as (4$\times$4) supercells of the surface unit cells in vacuum described above. They were separated by a water film with a diameter of about 60~\AA{} containing 576 molecules. In total, each studied interface system contains 4416 atoms. Before being placed in the vacuum between the slab surfaces, the aqueous regions were equilibrated in the $NVT$ ensemble for 25~ps by HDNNP-driven MD simulations at 300~K employing a density of 0.998~g/cm$^3$. The water films were then inserted in the vacuum regions of the slabs with an initial vertical separation of about 2.5~\AA{} between the water region and both surfaces. To close this gap, the water-belite interfaces were then equilibrated for 1~ns in the $NPT$ ensemble. After this $NPT$ simulation, for the central part of the water region with 25~\AA{} diameter, which is about 20~\AA{} away from both surfaces, a density of about 0.92 g/cm$^3$ was found, which is very close to the  RPBE+D3 equilibrium density of bulk water~\cite{P4556}. This is in agreement with previous studies on other solid-liquid interfaces showing that water films of this thickness adopt a bulk-like water structure in the center~\cite{P4859}. The properties of the interfaces were then determined in 1~ns MD simulations in the $NVT$ ensemble. 

\section{Radial distribution functions} 

\begin{figure}
  \centering
  \includegraphics[width=1.0\linewidth]{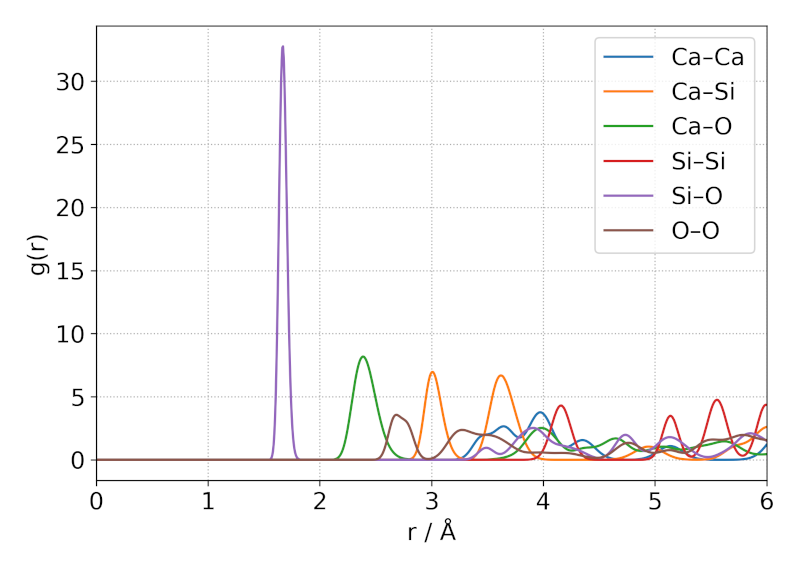}
  \caption{Radial distribution functions in bulk $\gamma$-belite obtained by MD simulations of a ($3\times 2\times 3$) supercell in the $NPT$ ensemble at 300~K.}
  \label{fig:rdf}
\end{figure}

Based on the HDNNP-driven \textit{NPT} MD simulation of the ($3\times 2\times 3$) supercell, the bulk $\gamma$-belite radial distribution function was calculated at 300~K (Fig~\ref{fig:rdf}). The strong first Si-O peak is related to the four oxygen atoms coordinating Si atoms in a tetrahedral arrangement. The first coordination shell of the calcium atoms can be found in a range from 2.1 to 2.9~\AA{} with a maximum at 2.4~\AA{}. 

\section{Reference data set construction} 

The reference data set consists of three parts: bulk $\gamma$-belite, bulk liquid water, and solid-liquid interface structures. The reference data set composition is compiled in Table~\ref{table:structural_informations}.

\begin{table}[!ht]
\centering
\caption{Composition of the reference data set. Given is the number of structures together with the respective number of atoms per structure for the three types of systems. }
\begin{tabular}{l r r} 
 \hline
  Type & Atoms & Structures \\
  \hline
Bulk $\gamma$-belite     &    &  \\
 & 28 & 2522 \\
       & 224  & 839 \\

Total       &  & 3361 \\
 \hline
Water     &    &  \\
     &    24 & 932 \\
     &    30 & 222 \\
     &    36 &  556 \\
     &    48 & 3106 \\
     &    96 & 900 \\
     &    192 & 920 \\
     &    384 & 120 \\
Total &   & 6756 \\
 \hline
Water-belite interfaces &    &  \\
     &    208 & 9013 \\
     &    304 & 4313 \\
     &    512 & 594 \\
     &   1620 & 347 \\
Total    &   & 14267 \\
\hline
Sum over all types & & 24384 \\
\hline
\end{tabular}
\label{table:structural_informations}
\end{table}

For bulk liquid water, geometries were taken from previous work~\cite{P4556}, recalculated with the DFT settings described above. The high-pressure ice structures in the dataset were not included in the present work. The initial bulk $\gamma$-belite structures were prepared based on random atomic displacements around the experimental and DFT-optimized geometries. In addition, the system was expanded and compressed by changing the unit cell parameters while keeping fractional coordinates of the atoms fixed. The initial reference data set of 971 geometries was used as a starting point for the iterative extension of the data set by an active learning approach - RuNNerActiveLearn \cite{Eckhoff2019,Eckhoff2021} - using MD simulations in the $NPT$ ensemble at a pressure of 1~bar and a temperature range of 100-500~K. The final reference data set contains 3361 bulk $\gamma$-Belite structures (2522 1$\times$1$\times$1 and 839 2$\times$2$\times$2 supercells). 

The third part of the reference data set contains interface structures of both investigated terminations as well as 32, 64, and 96 water molecules. Additionally, for T2 largers systems with 288 water molecules were prepared. Interfaces were constructed based on $\gamma$-belite slabs of different thicknesses. The initial interfacial structures were prepared with the following method:
first, periodic cells laterally matching the dimensions of the surface slabs containing 32 water molecules of density equal to 0.998~g/cm$^3$ were equilibrated through $NVT$ MD simulations at 300~K using a preliminary HDNNP for water trained to the bulk liquid water data only. Ten uncorrelated water structures were then inserted in the vacuum regions of optimized surfaces at a distance between 2.3-2.7~{\angstrom}. Prepared in this way, 50 interfaces were subsequently optimized using DFT with a force convergence criterion of 0.05~eV/{\angstrom}. From the obtained optimization pathways every 25th structure was included in the initial interface reference data set containing 522 structures for T3 and 513 for T2. The active learning procedure for the interfaces has been started for the most stable T3 termination, based on $NPT$ simulations at a  pressure of~1 bar along the $z$-direction perpendicular to the surface and a temperature range of 275-375~K. From this point, the HDNNP was trained based on a combined reference data set, containing bulk water, bulk $\gamma$-belite, and the interface structures. As soon as the MD simulations became stable for about 5-10~ps, the T2 termination of the (010) surface was also included in the active learning, which was then continued for both surfaces.  

For all types of geometries, only structures with absoluted forces smaller than 10 eV/{\angstrom} were included in the reference data set. 

\section{Atom-centered symmetry functions} 

The local atomic environments with a cutoff radius of 6~\AA{} are described by 188, 191, 167, and 165 atom-centered symmetry functions (ACSFs)~\cite{P2882} for hydrogen, oxygen, calcium, and silicon atoms, respectively.
Radial ACSF have been contructed for all atom pairs using the set of $\eta$ values 
0.00000000, 0.01396710, 0.06161784, and 0.48552766 bohr$^{-2}$. For Ca-Si, Ca-Ca, Si-Si, and Si-Ca, the symmetry function for the highest $\eta$ value has been omitted. Moreover, a second set of shifted radial symmetry functions with a maximum at 0.9 bohr and the $\eta$ values 1.0000, 3.5845, 6.1690, and 8.7535 bohr$^{-2}$ has been included. Here, the ACSF with $\eta$ equal to 1.0 bohr$^{-2}$ has been omitted for Ca-Si, Ca-Ca, Si-Si, and Si-Ca atom pairs. 
For the generation of an angular symmetry function, the starting point is a pool of combinations of $\lambda$=\{-1,1\}, $\zeta$=\{1,2,4,8\}, and $\eta$=\{0,0.06161784\}. From this pool the following ACSFs were removed for specific element triples, as the range of values has been small:

\begin{tabular}{l l l l} 
Element triple & $\eta$ & $\lambda$ & $\zeta$ \\
H-Si-Si  & 0.06161784 & -1.0 & 8.0 \\
H-Si-Si  & 0.06161784 &  1.0 & 2.0, 4,0 , 8.0 \\
O-Si-Si  & 0.06161784 &  1.0 & 8.0 \\
Si-H-Si  & 0.06161784 & -1.0 & 1.0, 2.0, 4.0, 8.0 \\
Si-O-Si  & 0.06161784 & -1.0 & 4.0, 8.0 \\
Si-O-Si  & 0.00000000 & -1.0 & 4.0, 8.0 \\
Si-Si-Si & 0.00000000 & -1.0 & 4.0, 8.0 \\
Si-Si-Si & 0.06161784 & -1.0 & 1.0, 2.0, 4.0, 8.0 \\
Si-Si-Si & 0.06161784 &  1.0 & 1.0, 2.0, 4.0, 8.0 \\ 
Si-Si-Ca & 0.06161784 & -1.0 & 1.0, 2.0, 4.0, 8.0 \\
Si-Si-Ca & 0.06161784 &  1.0 & 8.0 \\
Si-Ca-Ca & 0.06161784 & -1.0 &  4.0, 8.0 \\ 
Ca-H-Si  & 0.06161784 & -1.0 & 8.0 \\
Ca-H-Ca  & 0.06161784 & -1.0 &  4.0, 8.0 \\
Ca-O-Si  & 0.06161784 & -1.0 & 8.0 \\ 
Ca-O-Ca  & 0.06161784 & -1.0 & 4.0, 8.0 \\
Ca-Si-Si & 0.06161784 &  1.0 & 2.0, 4.0, 8.0 \\
Ca-Si-Ca & 0.06161784 & -1.0 & 4.0, 8.0 \\
Ca-Ca-Ca & 0.06161784 & -1.0 & 1.0, 2.0, 4.0, 8.0 \\
Ca-Ca-Ca & 0.06161784 &  1.0 &  4.0, 8.0 \\
\end{tabular}

\section{$RuNNer$ settings} 

The input settings used for training HDNNPs with $RuNNer$ are listed in Table~\ref{table:RuNNer_settings}. The different HDNNPs used for active learning  have been obtained by using different random numbers for the initial neural network weights as well as the splitting into a training and test set. 

\begin{table}[!ht]
\centering
\caption{Settings of the $RuNNer$ input file used for training the HDNNPs.}
\begin{tabular}{l l} 
 \hline
  \textbf{Keyword} & \textbf{Value}  \\
  \hline
  nn\textunderscore type\textunderscore short & 1 \\
  random\textunderscore number\textunderscore type & 5 \\
  random\textunderscore seed & 316 \\
  number\textunderscore of\textunderscore elements & 4 \\
  elements & H O Si Ca \\
  cutoff\textunderscore type & 1 \\
  use\textunderscore short\textunderscore nn & \\
  global\textunderscore hidden\textunderscore layers\textunderscore short & 2 \\
  global\textunderscore nodes\textunderscore short & 15 15  \\
  global\textunderscore activation\textunderscore short & ttl \\
  test\textunderscore fraction & 0.1 \\
  epochs & 30 \\
  repeated\textunderscore energy\textunderscore update & \\
  mix\textunderscore all\textunderscore points & \\
  scale\textunderscore symmetry\textunderscore functions & \\
  center\textunderscore symmetry\textunderscore functions & \\
  fitting\textunderscore unit & eV \\
  precondition\textunderscore weights & \\
  use\textunderscore short\textunderscore forces & \\
  kalman\textunderscore lamda\textunderscore short & 0.98 \\
  kalman\textunderscore nue\textunderscore short & 0.9987 \\
  short\textunderscore energy\textunderscore fraction & 0.01\\
  short\textunderscore force\textunderscore fraction & 1.0 \\
  force\textunderscore scale\textunderscore update & 4.0 \\
  nguyen\textunderscore widrow\textunderscore weights\textunderscore short & \\
  short\textunderscore force\textunderscore error\textunderscore threshold & 1.0 \\
  short\textunderscore energy\textunderscore error\textunderscore threshold & 0.1 \\
  remove\textunderscore atom\textunderscore energies & \\
  atom\textunderscore energy                  &  Ca -682.06533 \\
  atom\textunderscore energy                    & Si -290.32622\\
  atom\textunderscore energy                    & O -75.18080 \\
  atom\textunderscore energy                    & H -0.45891 \\
  \hline
\end{tabular}
\label{table:RuNNer_settings}
\end{table}

\section{Energy and force error plots} 
\newpage

\begin{sidewaysfigure}
  \centering
  \includegraphics[width=1.0\linewidth]{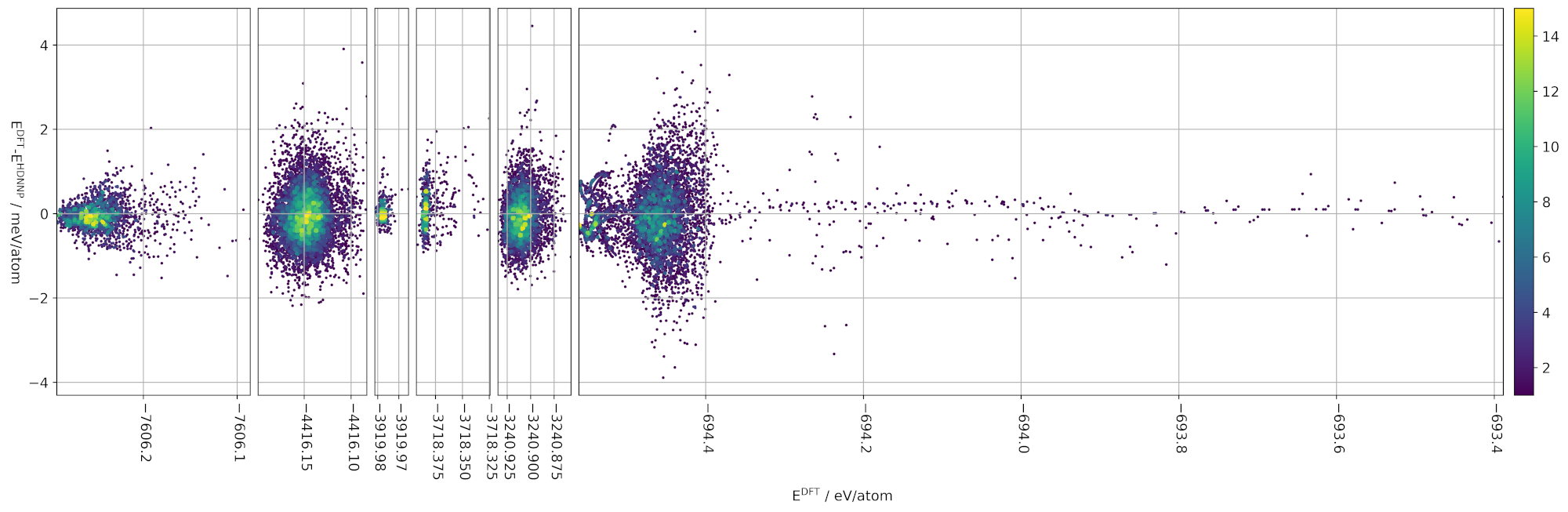}
  \caption{Energy error plots showing the deviations between the HDNNP predictions and the DFT reference values for the training data set. The data points are colored based on their relative density, highlighting regions of higher data population.}
  \label{fig:energy-train}
\end{sidewaysfigure}
    
\begin{sidewaysfigure}
  \centering
  \includegraphics[width=1.0\linewidth]{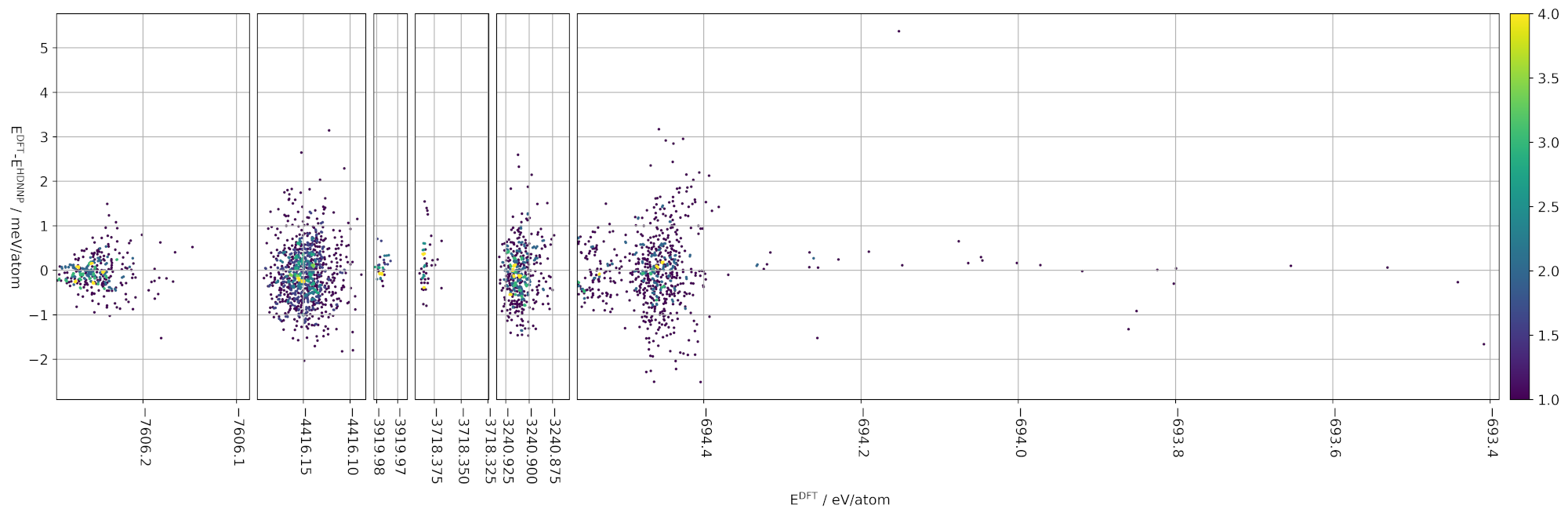}
  \caption{Energy error plots showing the deviations between the HDNNP predictions and the DFT reference values for the test data set. The data points are colored based on their relative density, highlighting regions of higher data population.}
  \label{fig:energy-test}
\end{sidewaysfigure}

\newpage

\begin{figure*}[!ht]
  \centering
  \includegraphics[width=0.84\linewidth]{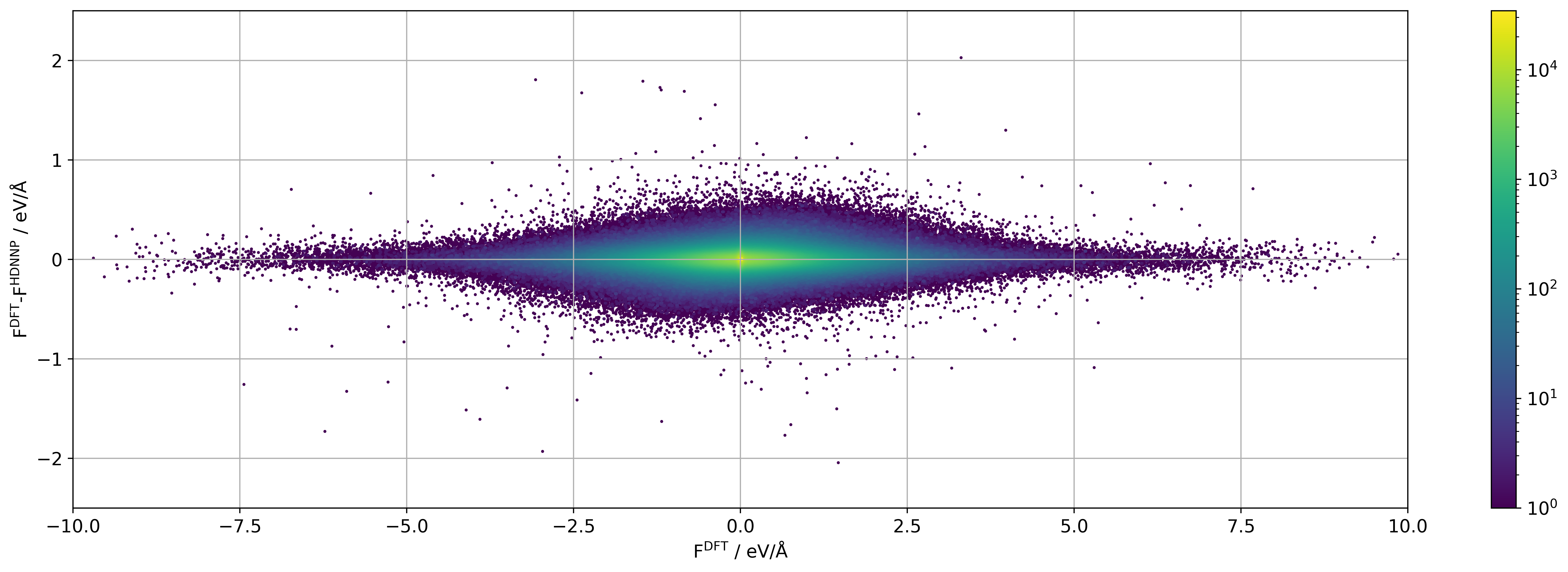}
  \caption{Force error plots showing the deviations between the HDNNP predictions and the DFT reference values for the training data set. The data points are colored based on their relative density, highlighting regions of higher data population.}
  \label{fig:force-train}
\end{figure*}

\begin{figure*}[!ht]
  \centering
  \includegraphics[width=0.84\linewidth]{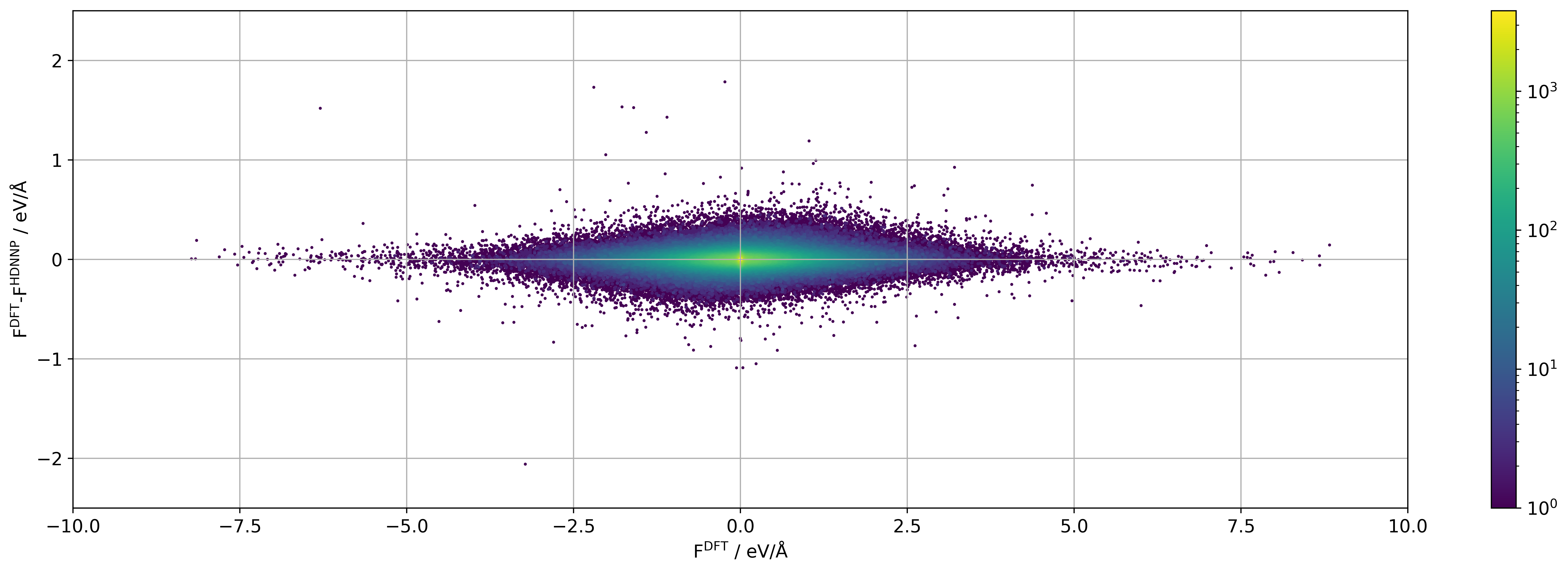}
  \caption{Force error plots showing the deviations between the HDNNP predictions and the DFT reference values for the test data set. The data points are colored based on their relative density, highlighting regions of higher data population.}
  \label{fig:force-test}
\end{figure*}

\begin{figure}[!ht]
  \centering
  \includegraphics[width=0.92\linewidth]{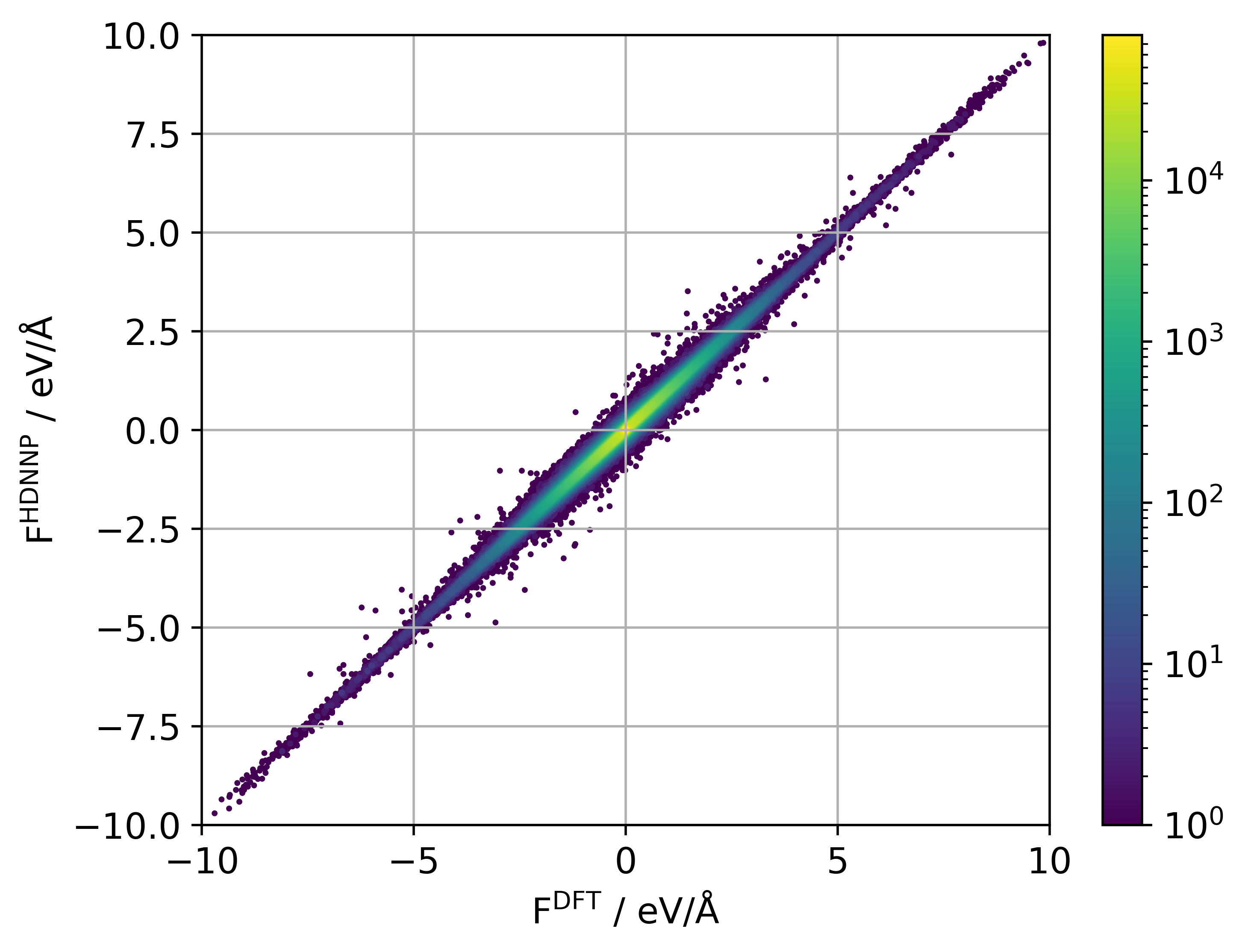}
  \caption{Force error plots showing the correlation between the HDNNP predictions and the DFT reference values for the training data set. The data points are colored based on their relative density, highlighting regions of higher data population.}
  \label{fig:force-train-cor}
\end{figure}

\begin{figure}[!ht]
  \centering
  \includegraphics[width=0.92\linewidth]{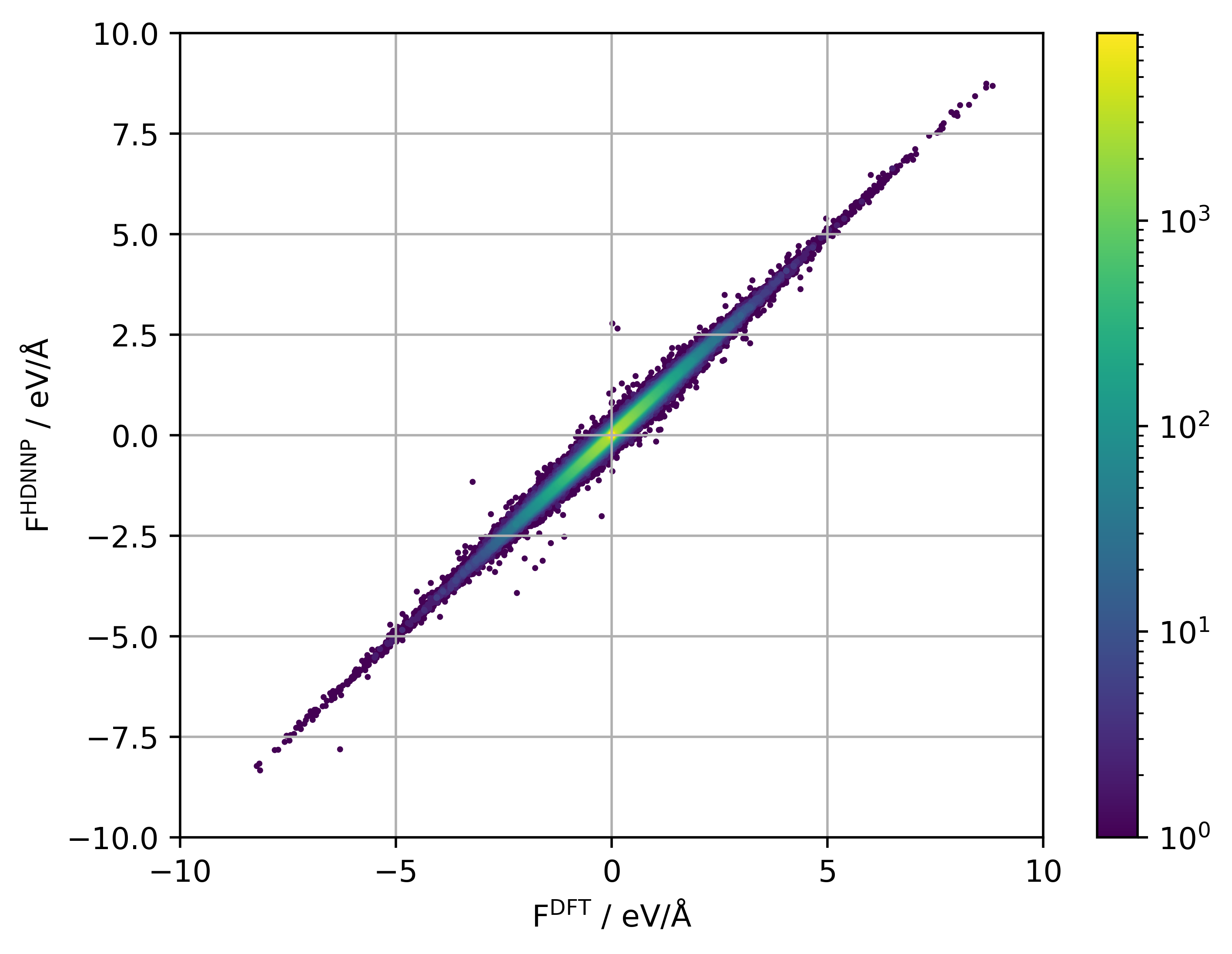}
  \caption{Force error plots showing the correlation between the HDNNP predictions and the DFT reference values for the test data set. The data points are colored based on their relative density, highlighting regions of higher data population.}
  \label{fig:force-test-cor}
\end{figure}

\newpage
Figures~\ref{fig:energy-train} and~\ref{fig:energy-test} present the deviations between the HDNNP and the DFT-D3 reference energies as a function of the reference energy for the training and test data sets, respectively. Plots are split in multiple segments due to the large offsets in total energies for different parts of the reference data set due to different stoichiometries.  For both data sets the difference between the predicted and reference energies for the majority of structures is smaller than 1~meV/atom. 

Figures~\ref{fig:force-train} and~\ref{fig:force-train} present the force component deviations for the training and test data sets. For forces in the range from -2.5 to 2.5 eV/~\AA{} very rarely differences between the HDNNP and reference forces of up to 1~eV/~\AA{} are observed. However, in the full reference data set containing 14,453,784 force components, only 1.5\% of the force components have an absolute value of force deviations larger than the force 3*RMSE of the test set (3*RMSE(F) = 0.2052 eV/\AA{}). Additionally, Figures~\ref{fig:force-train-cor} and~\ref{fig:force-test-cor} present the correlation between the HDNNP-predicted force components and their reference values.

\section{Equilibration of the MD simulations}
\begin{figure*}[ht]
  \centering
  \includegraphics[width=1.0\linewidth]{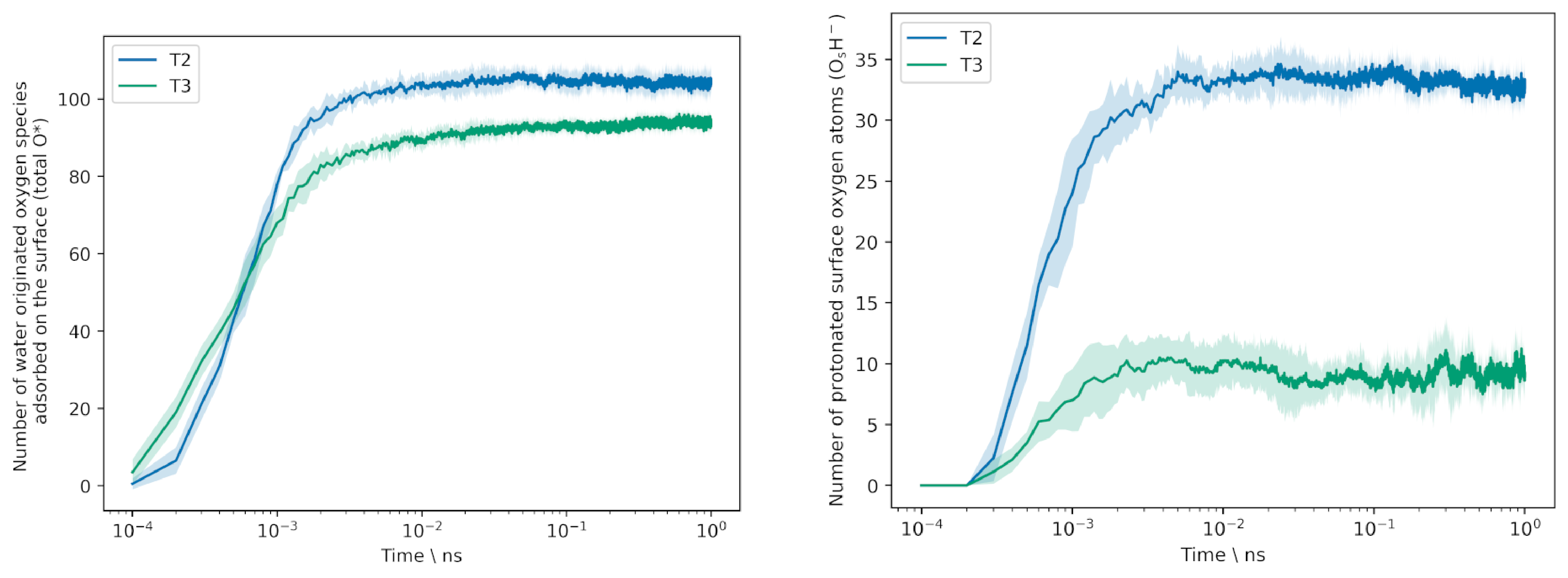}
  \caption{Number of average adsorbed oxygen atoms (a) and protonated surface oxygen atoms (b) during \textit{NPT} MD simulations at 300~K of the T2 and T3 terminations, plotted with standard deviation over 8 trajectories.}
  \label{fig:NPT1}
\end{figure*}

\begin{figure*}[ht]
\centering
  \includegraphics[width=1.0\linewidth]{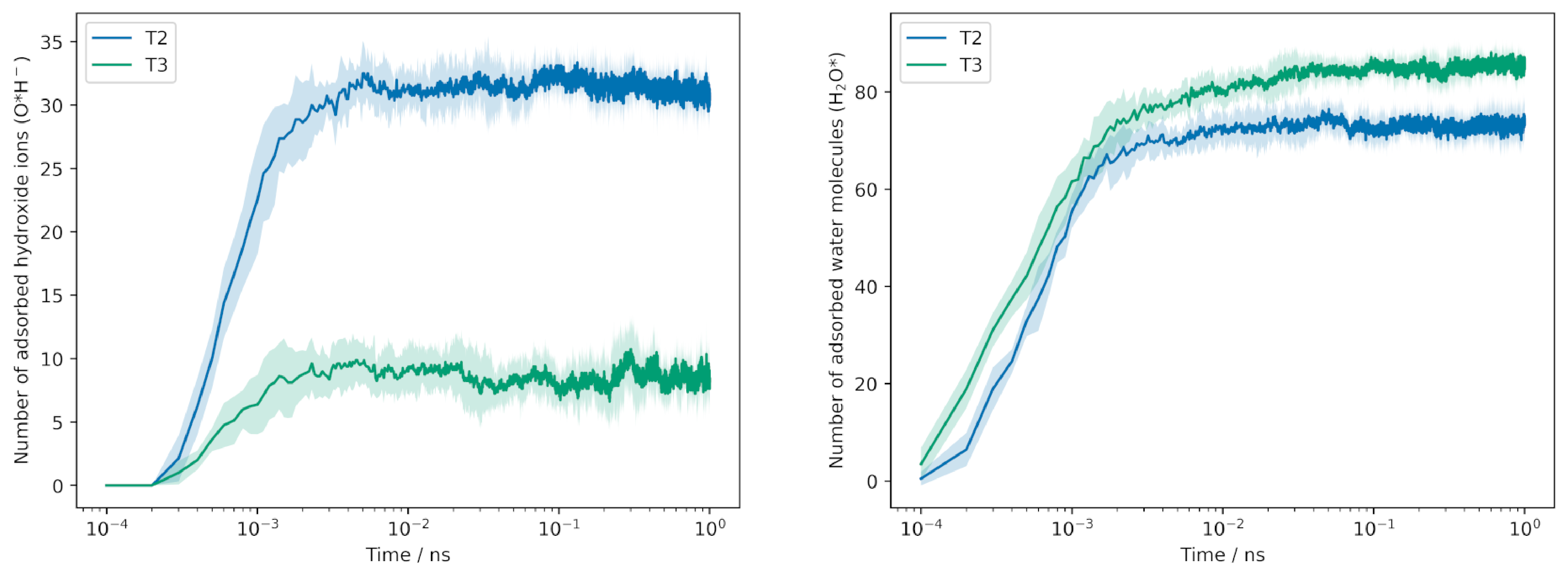}
\caption{Number of average adsorbed oxygen atoms with distinction between hydroxide ions (a) and  water molecules (b) during \textit{NPT} MD simulations at 300~K of the T2 and T3 terminations, plotted with standard deviation over 8 trajectories.}
\label{fig:NPT2}
\end{figure*}

Figures~\ref{fig:NPT1} and~\ref{fig:NPT2} show the time evolution of different oxygen species at T2 and T3 during the initial equilibration of the system by MD in the $NPT$ ensemble at 300~K. After about 1~ns simulation time equilibrium has been reached.

\end{appendices}
\clearpage
\bibliography{Literature}


\end{document}